\documentclass[conference]{IEEEtran}
\usepackage{cite}
\usepackage{amsmath,amssymb,amsfonts}
\usepackage{graphicx}
\usepackage{textcomp}
\usepackage{cuted}
\usepackage[usenames,dvipsnames]{xcolor}

\usepackage[utf8]{inputenc} 
\usepackage[T1]{fontenc}    
\usepackage{hyperref}       
\usepackage{url}            
\usepackage{booktabs}       
\usepackage{nicefrac}       
\usepackage{microtype}      
\usepackage{lipsum}
\usepackage{float}
\usepackage{braket}
\usepackage{enumitem}

\usepackage[caption=false,font=footnotesize]{subfig}
\graphicspath{{figures/}}

\def\BibTeX{{\rm B\kern-.05em{\sc i\kern-.025em b}\kern-.08em
    T\kern-.1667em\lower.7ex\hbox{E}\kern-.125emX}}

\begin{document}

\title{Characterizing the Stability of NISQ Devices}

\author{\IEEEauthorblockN{Samudra~Dasgupta}
\IEEEauthorblockA{\textit{Quantum Computing Institute} \\
\textit{Oak Ridge National Laboratory}\\
Oak Ridge, Tennessee, USA \\
dasguptas@ornl.gov}
\and
\IEEEauthorblockN{Travis S.~Humble}
\IEEEauthorblockA{\textit{Quantum Computing Institute} \\
\textit{Oak Ridge National Laboratory}\\
Oak Ridge, Tennessee, USA \\
humblets@ornl.gov}
}

\maketitle

\begin{abstract}
In this study, we focus on the question of stability of NISQ devices. The parameters that define the device stability profile are motivated by the work of DiVincenzo in \cite{divincenzo} where the requirements for physical implementation of quantum computing are discussed. We develop the metrics and theoretical framework to quantify the DiVincenzo requirements and study the stability of those key metrics. The basis of our assessment is histogram similarity (in time and space). For identical experiments, devices which produce reproducible histograms in time, and similar histograms in space, are considered more reliable. To investigate such reliability concerns robustly, we propose a moment-based distance (MBD) metric. We illustrate our methodology  using data collected from IBM's Yorktown device. Two types of assessments are discussed: spatial stability and temporal stability.
\end{abstract}

\begin{IEEEkeywords}
Quantum Computing, Benchmarks and Metrics, NISQ Stability, Moment Based Distance (MBD), Histogram Similarity
\end{IEEEkeywords}

\section{Introduction}
Quantum computing is a model of computation built on the principles of quantum mechanics that could enable fundamental breakthroughs in diverse fields such as chemistry, AI, finance, communications and internet security. Building quantum computers that realize these principles requires high-fidelity control over quantum physical systems and their interactions, and a variety of different physical systems are now being developed to demonstrate these ideas. Leading examples of early devices using superconducting electronics, trapped ions, photonics, and silicon among many others provide important insights needed for the design and operation of future quantum computing systems. 

While current devices fail to meet the requirements for fault-tolerant quantum computing \cite{gottesman1998}, they are an important milestone towards that goal. In particular, noisy intermediate-scale quantum (NISQ) devices offer an exciting frontier for testing how the principles of quantum computing may be used in practice under experimental conditions \cite{preskill2018}. By controlling quantum mechanical phenomena, such as superposition and randomness, NISQ devices provide first-in-kind platforms to demonstrate the control and characterization of quantum physical systems with computational logic \cite{kandala2017hardware,dumitrescu2018cloud,hempel2018quantum,klco2018quantum,roggero2020quantum}.
Key results using NISQ devices include characterization, verification, and validation, testing and evaluation of quantum computer programs, and performance assessments of quantum applications. Recent demonstration of so-called quantum supremacy using NISQ devices \cite{arute2019} is a major milestone towards quantum computational advantage.

Fundamental to the observations taken from experimental NISQ devices is the statistical significance of the reported results, which is required to differentiate signals of interest from noisy background processes. As the repeated use of a specific NISQ device is well understood to generate a series of noisy results, statistical significance is typically measured with respect to the expectation value or arithmetic mean of the observed outcomes relative to the corresponding standard deviation. By evaluating these metrics against the expected outcomes, decisions and assessments of confidence in these results can be made. A central pillar to the interpretation of statistical significance is the expectation that the underlying NISQ device is \textit{reliable}, which we define below in terms of the stability of the generated noisy distribution of outcomes. If lacking reliability, usage of NISQ devices raises questions as to whether empirical results derived from such a device are reproducible, a critical tenet of the modern scientific method. 

Here we address the characterization of stability of NISQ devices using statistical tests that quantify variation in device metrics across varying time periods. Our approach is based on the comparison on experimental observation across both space and time to identify when the underlying probabilistic processes change. Our approach is based on the assessment of NISQ reliability in terms of the common criteria upon which these devices are designed and operated. As originally described by DiVincenzo, quantum computing devices must meet a set of criteria to support (digital) quantum computation \cite{divincenzo}. The list of criteria is
\begin{itemize}
    \item Host a scalable array of well-defined and addressable quantum physical systems
    \item Support initialization of the quantum state of these physical systems to a well-define fiducial state
    \item Realize a universal set of gate operations
    \item Maintain the coherence of the quantum state much longer than the duration of any individual gate
    \item Support addressable measurements of each quantum physical system
\end{itemize}
Each of these design criteria may be quantified by a corresponding metric and we present a working set of metrics for the DiVincenzo criteria in Table~\ref{tab:dvz} alongside their representative symbols and definitions.
\begin{table}[ht]
    \caption{Device metrics for assessing the DiVincenzo design criteria}
    \label{tab:dvz}
    \centering
    \begin{tabular}{|p{3.1cm}|p{4.4cm}|}
        \hline 
        \textit{Device Metric} (\textit{Symbol}) & \textit{Description} \\ \hline
        Register Capacity ($n$) & The maximal amount of information that may be stored in the register.  \\ \hline
        Initialization Fidelity ($F_{\textrm{I}}$) & The accuracy with which the register state is prepared as the fidelity state. \\ \hline
        Gate Fidelity ($F_{\textrm{G}}$) & The accuracy with which a gate operation prepares the expected output. \\ \hline
        Duty Cycle ($\tau$) & The ratio of the gate duration to coherence time. \\ \hline
        Addressability ($F_{\textrm{A}}$) & The ability to address qubits individually. \\ \hline
    \end{tabular}
\end{table}

A variety of methods exist for estimating the device metrics presented in Table~\ref{tab:dvz}, but there is a dearth of understanding of how to assess the reliability of these metrics. In particular, for NISQ devices, the underlying noise processes are poorly understood and efforts to assert statistical significance are complicated by questions of the reliability in device performance. That is to say, whereas individual estimates of device metrics may be statistically significant at the time they are acquired, those estimates may be unreliable descriptions of the device at later times due to fluctuations in the underlying noise. The current lack of stability in estimates of device metrics poses a problem for reproducing key results. Indeed, this raises the concern of a reproducibility crisis for a large variety of studies currently undertaken by the quantum computing community using NISQ devices.
\par
This paper is organized as follows:
Section~\ref{mbd} presents a framework by which the reliability of noisy metrics is assessed in terms of the stability of the underlying statistical distribution. This yields a reliability metric defined by the moment-based distance between two distributions, for which we discuss the mathematical properties of the metric and study its characteristics through numerical simulation. In Sec.~\ref{stability}, we apply the moment-based distance to evaluate the spatial and temporal stability of NISQ devices relative to the DiVincenzo design criteria. We illustrate the subsequent stability assessment using the five-qubit transmon device from IBM called Yorktown. We present summary conclusion in Sec.~\ref{conc}.
\section{Measuring Statistical Stability}\label{mbd}
The repeated use of a NISQ device to execute a quantum circuit often generates widely varying outcome distributions, even when the arithmetic means of these distributions agree. Expectation values are a coarse measure by which to compare noisy processes and do not directly address the question of device reliability, which we define as the ability to reproduce the generated probability distributions. In this section, we introduce a robust distance metric for measuring NISQ reliability in terms of the similarity between statistical distributions. We show that this relates to the concept of stability for distributions that may be used to test whether a device is stable.
\subsection{Theory of Moment-Based Distance (MBD)}
A metric to benchmark distribution reproducibility should be simple, intuitive and yet mathematically robust. Contemporary research in quantum computing often use Fidelity and Kullback–Leibler divergence as distribution distance measures (see \cite{nielsen}). However, neither Fidelity nor Kullback–Leibler divergence satisfy the triangle inequality and hence do not qualify as a \textit{true distance metric}. Such \textit{non-metric} distance measures can be misleading when used for benchmarking. Another measure commonly in use is the Total Variation Distance (TVD). While it is a true distance metric, it ignores higher order effects (such as kurtosis and skew). This motivated us to develop a moment-based approach to distribution distance (we provide a side-by-side comparison of our approach to TVD later).
\par
If two distributions are identical then all their moments are equal. Conversely, if all the moments are equal then two distributions are identical almost always. There are some pathological cases for which the converse does not hold true but they are artificial, corner cases which do not arise in real-life.
\par
This motivates our definition of a non-euclidean moment-based distance metric $d$ between two histograms $f$ and $g$ as
\begin{equation}
d(f,g) = \sum\limits_{m=0}^{\infty} S_m(f, g)
\end{equation}
where
\begin{equation}
S_{m}(f,g) = \frac{1}{(m)!}\int\limits_a^{b} \left| \left( \frac{x}{\gamma} \right)^{m}(f(x) - g(x))\right|dx
\end{equation}
and,
\begin{equation}
\gamma = \max( |a|,|b|)\\
\end{equation}
for a bounded real variable $x$. Here, $a$ and $b$ are the minimum and maximum values of $x$ and can be derived from theoretical considerations (e.g., when the random variable is a probability then $\gamma=1$) or from empirical histogram data.
\par
The moment-based-distance $d(f,g)$ satisfies the following properties:
\begin{enumerate}
\item $d(f,g) \geq 0$ follows from the definition of $d$.

\item $d(f, g) = d(g, f)$ follows from the definition of $d$.

\item $d(f, g) = 0$ iff $f(x) = g(x)$.\\
Proof: If $f(x) = g(x)$, then $d = 0$ because $S_m = 0$ for every $m$. Conversely, if $d = 0$, then $S_{m} = 0$ for all $m$.If $S_{m} = 0$, then 
for all $x$, the integrand  must satisfy
\begin{equation*}
\left| \left(\frac{x}{\gamma}\right)^m (f-g)\right| = 0
\end{equation*}
As $(x)^{m} \neq 0$ for all $x$, it must be that $|f(x)-g(x)| = 0$ for all $x$ and, hence, $f(x) = g(x)$.
\hfill $\tiny\blacksquare$

\item $d(f,g) \leq d(f, h) + d(h, g)$ \\
Proof: For every $m$,
\begin{equation*}
\begin{split}
S_m(f,g) =& \int\limits_a^{b} \left| \left( \frac{x}{\gamma} \right)^{m}\frac{1}{m!}(f(x) - g(x))\right|dx\\
=& \int\limits_a^{b} \bigg| \left( \frac{x}{\gamma} \right)^{m}\frac{1}{m!}(f(x) - h(x) \\
& \hspace{2.0cm}  + h(x) - g(x))\bigg|dx\\
\leq & \int\limits_a^{b} \left| \left( \frac{x}{\gamma} \right)^{m}\frac{f(x) - h(x)}{m!}\right|dx\\
& +\int\limits_a^{b} \left| \left( \frac{x}{\gamma} \right)^{m}\frac{h(x) - g(x)}{m!}\right|dx\\
\leq& S_m(f,h) + S_m(h,g)\\
\end{split}
\end{equation*}
and whence the sum satisfies the inequality as well.\hfill $\blacksquare$

\item The series $d = S_0 + S_1 + S_2 + \cdots$ converges.\\
Proof: The distance $d$ converges if, after some fixed term, the ratio of each term to the preceding term is less than some quantity $r$, which is itself numerically less than unity \cite{hallnknight}. If $S_{m+1} < S_m$ for all $m \geq 1$, then
\begin{equation*}
\begin{split}
d =& S_0 + S_1 + S_1 \cdot \frac{S_2}{S_1} + S_1\cdot \frac{S_2}{S_1}\cdot \frac{S_3}{S_2} + \cdots\\
<& S_0 + S_1(1+r+r^2 +r^3 + \cdots)\\
=& S_0 + \frac{S_1}{1-r}  \textrm{ since } r<1\\
\end{split}
\end{equation*}
To prove that $S_{m+1} < r S_m$ for $m\geq 1$, we proceed as follows:
\begin{equation*}
\begin{split}
S_{m+1} &= \int\limits_a^{b} \left| \left( \frac{x}{\gamma} \right)^{m+1}\frac{1}{(m+1)!}(f(x) - g(x))\right|dx \\
&= \int\limits_a^{b} \left| \frac{x}{\gamma(m+1)}  \right|\left|\left( \frac{x}{\gamma} \right)^{m}\frac{f(x) - g(x)}{m!}\right|dx\\
&\leq \left| \frac{x}{\gamma} \right|_{max} \frac{1}{m+1} S_m\\
&\leq \frac{1}{m+1}S_m \textrm{ since }  \left| \frac{x}{\gamma} \right|_{max} = 1 \\
&\leq \frac{1}{2}S_m \textrm{ since } m \geq 1
\end{split} 
\end{equation*}
where $|x|_{max}$ is the maximum of $x$. \hfill $\blacksquare$
\end{enumerate}
\par
An important consequence of the latter convergence property is that the moment-based distance satisfies the practical requirement that lower-order moments contribute more than higher-order moments to the distance (for $m>1$). This proves essential to our subsequent use of the moment-based distance below, as we rely on the approximate distance defined to order $n$ as
\begin{equation}
    d_{n} = \sum_{m=0}^{n}{S_m}
\end{equation}
\subsection{Simulation Studies}
We next present a series of simulation studies to develop intuition for how the moment-based distance behaves in the presence of both stable and unstable distributions. In particular, we will show that moment-based distance is small but non-zero for distributions that are similar but not identical, while such deviations grow with dissimilarity.
\par
For our studies, we computed the distance of 10 different distributions with respect to a reference distribution. Table~\ref{table:dis_distance} summaries the list of tests as well as their moment-based distance from the reference normal distribution $N(\mu, \sigma)$. For testing purpose, the parameters are $\mu = 0.4, \Delta = 0.2$ and $\sigma = 0.04$. We note that, as expected, the distribution `closest' to $N(\mu, \sigma)$ is $N(1.01\mu, \sigma)$ and the `farthest' are $N(2\mu, \sigma)$ and $N(\mu+2\Delta, 2\sigma)$.
\begin{table}[htbp]
\renewcommand{\arraystretch}{1.3}
\caption{Moment-based distance by Distribution}
\label{table:dis_distance}
\centering
\begin{tabular}{|l|c|c|c|}
\hline
Distribution & $d_4$ & $d_{20}$ & Error(\%)\\
\hline
$N(\mu, \sigma)$ & 0.00000 & 0.00000 & NA\\
\hline
$N(\mu+\Delta, \sigma)$ & 2.70868 & 2.70876 & -0.00289\\
\hline
$N(\mu, 2\sigma)$ & 0.83252 & 0.83253 & -0.00104\\
\hline
$N(\mu, 4\sigma)$ & 1.47301 & 1.47304 & -0.00180\\
\hline
$N(2\mu, \sigma)$ & 2.93489 & 2.93520 & -0.01033\\
\hline
$N(\mu, 1.5\sigma)$ & 0.49215 & 0.49216 & -0.00091\\
\hline
$N(1.01\mu, \sigma)$ & 0.11739 & 0.11740 & -0.00079\\
\hline
$Skewed Normal(\mu, 2\sigma)$ & 0.80887 & 0.80888 & -0.00140\\
\hline
$Gumbel(\mu, 2\sigma)$ & 0.95131 & 0.95134 & -0.00246\\
\hline
\end{tabular}
\end{table}
\par
\begin{figure}[htbp]
\centering
\includegraphics[width=\columnwidth]{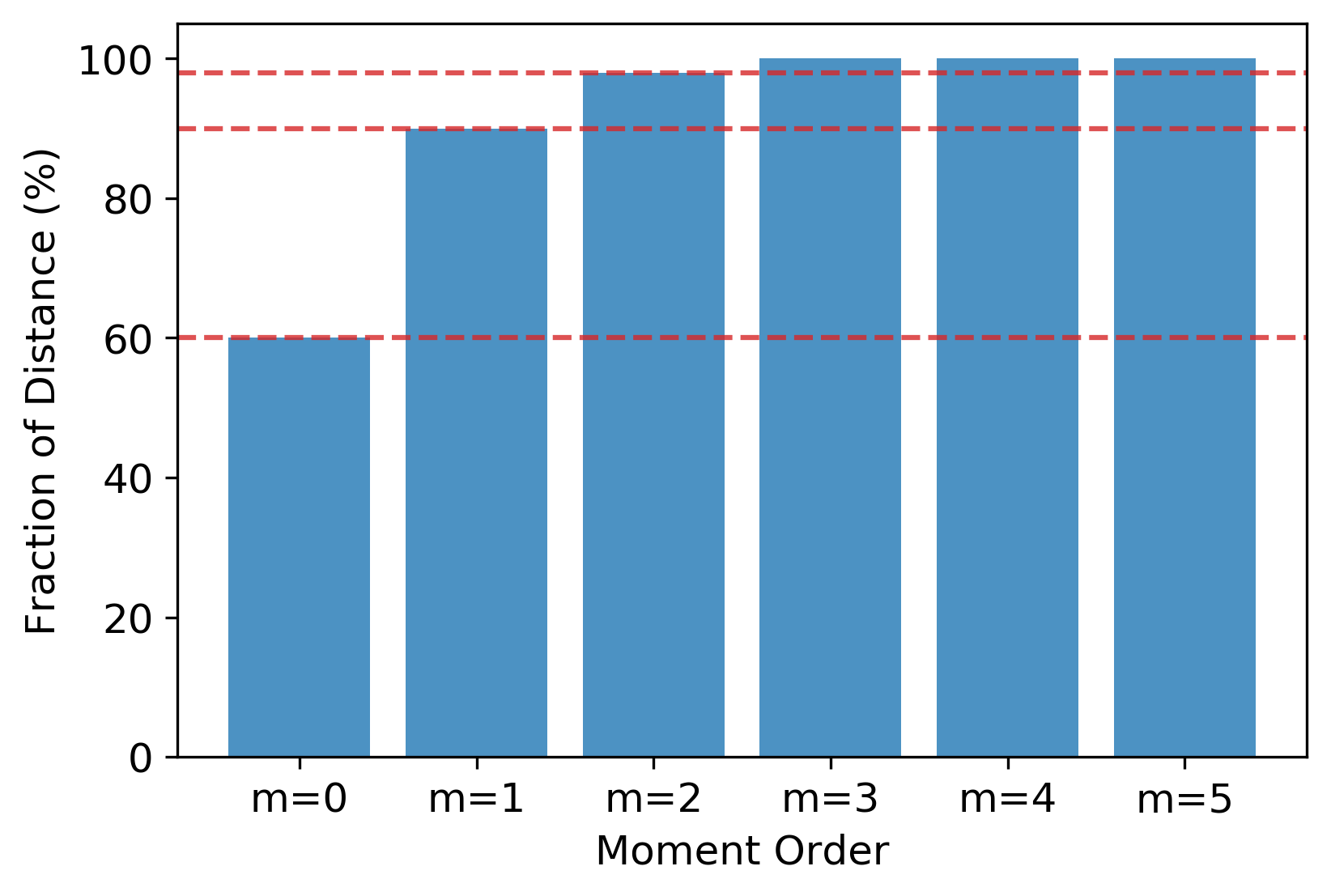}
\caption{Contribution to moment based distance ($d$) from increasing moment orders. The graph shows the results of comparing two normal distributions: $\mathcal{N}_1(\mu = \mu_0, \sigma=\sigma_0)$ and $\mathcal{N}_2(\mu = 2\mu_0, \sigma=2\sigma_0)$ where $\mu_0 = 40$ and $\sigma_0=4$.}
\label{fig:dnN_unequal_relative}
\end{figure}

We next study how the order of the series $d_n$ increases the accuracy of the distance measured. In our simulation studies of well-defined distributions, we find that $d_n$ converges for $n = 4$ when the distributions are sufficiently dissimilar. As shown in Fig. ~\ref{fig:dnN_unequal_relative}, the relative contributions of each $S_m$ to $d_n$ decreases with increasing $m$ as expected from the convergence property. Thus, $m=0$ accounts for about $60\%$ of the total distance while $m=1$ accounts for $90\%$ and $m=2$ reaches $98\%$. For $m=4$, $d_{m}$ is nearly $100\%$ of the $d_\infty$. Consequently, we will consider $m=4$ sufficient to accurately characterize the moment-based distance for the remainder of our analysis. This is certainly an approximation in the sense that two histograms which start to differ only after the fourth order moment will be erroneously classified as same. Is $d_4$ still a valid \textit{distance metric}? Yes. A glance at the proofs (in Section II-A) will reveal that properties (1) to (4) are still satisfied when we truncate the d series at a finite m (say m=4). Moreover, it converges too (i.e. Property (5) is satisfied too) because a finite number of terms (in this case 5 terms) is by definition convergent when the individual terms are finite. The latter is true because each $S_m$ is bounded between finite $a$ and $b$ as per Equation (2).
\par
\begin{figure}[htbp]
\centering
\includegraphics[width=\columnwidth]{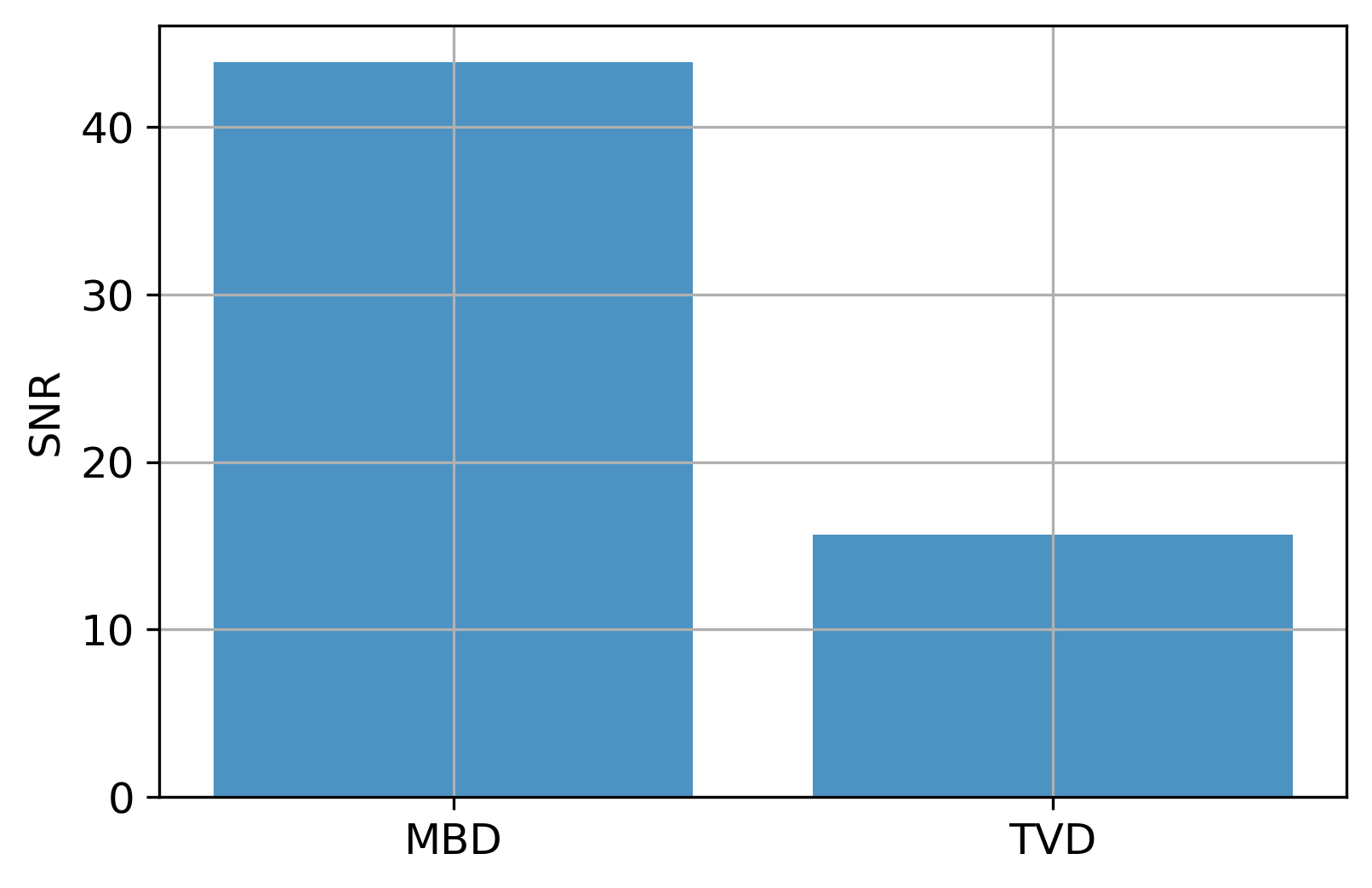}
\caption{Simulated signal-to-noise ratios of the moment-based distance and total variation distance for two normal distributions of varying width.}
\label{fig:MBDstability}
\end{figure}

As a point of comparison, we contrast the moment-based distance to total variation distance (TVD), a state-of-the-art metric which has proven useful in earlier experimental investigations \cite{buadescu2019quantum,rudinger2019probing}. We note that the magnitudes of the moment-based distance and total variation distance are not directly comparable as they follow very different methodologies but one can compare the corresponding signal-to-noise ratio (SNR) of the two metrics as the inverse of the coefficient of variation. For our numerical studies, we generated two time series, each of length 8192, by sampling two different probability distributions. The first was a normal distribution with mean 10 and standard deviation 1, and the second a normal distribution with mean 10 and standard deviation 4. We calculate the moment-based distance and total variation distance between these two time series, and then we repeated this numerical experiment 400 times to generate a distribution of the TVD and MBD distances. Using the average $\mu$ and standard deviation $\sigma$ of these distributions, we calculated the respective SNR as
\begin{equation}
\textrm{SNR} = \frac{\mu}{\sigma}
\end{equation}
As shown in Fig.~\ref{fig:MBDstability}, our results indicate that the moment-based distance has more statistical power as indicated by a higher SNR. 
\par
\begin{figure}[htbp]
\centering
\includegraphics[width=\columnwidth]{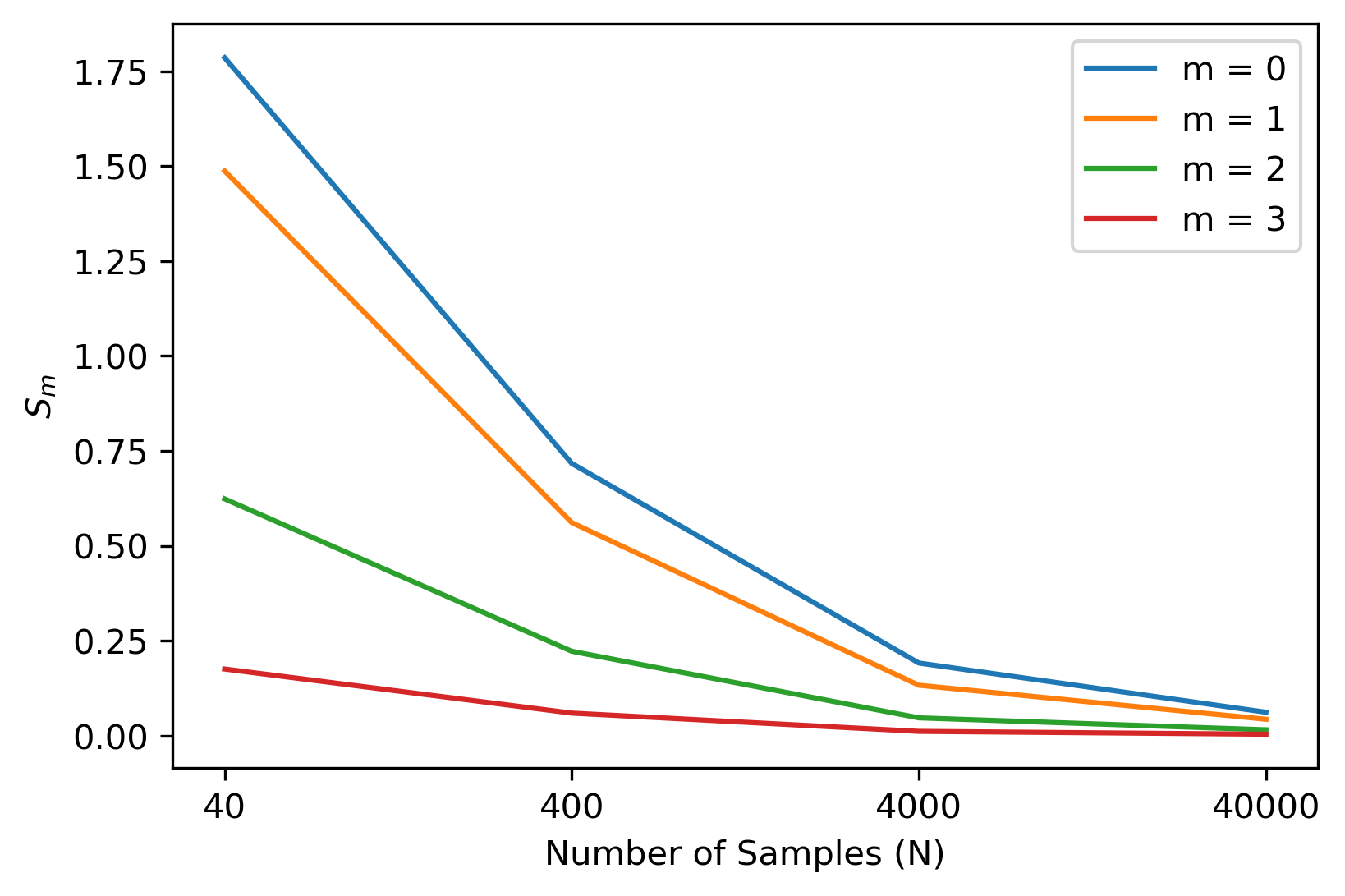}
\caption{When the two distributions are similar, then we expect each $S_m$
to be zero. Empirically, that happens as we increase the sample size. The
lower order moments take longer to go to zero.}
\label{fig:SnN_equal}
\end{figure}

As an aside, a practical concern is the dependence of precision of the moment-based distance on sampling size. Although each $S_m$ should vanish when two distributions are similar, finite sampling lead to approximations and ultimately a non-zero distance. As shown in Fig.~\ref{fig:SnN_equal}, increasing sampling may be used to reduce the relative error in each moment to a desired relative precision.
\section{Quantifying Device Reliability}\label{stability}
We define device reliability with respect to the stability of the statistical distributions characterizing the device properties in both space and time. Stability is, in turn, defined as the similarity between the experimentally observed distributions as measured by the moment-based distance. For example, as outlined below, we assess the temporal stability of device properties by evaluating the moment-based distance for a selected metric at regular temporal intervals. Similarly, we assess spatial stability in terms of the moment-based distance between observed distributions generated at different locations within the device, e.g. by different qubits. In both cases, decisions regarding reliability are framed as statistical process monitoring using the moment-based distance to assess stability and ultimately, control. For example, the moment-based distance may be used to define a Shewart control chart, by which a stability threshold is chosen to decide if the device is sufficiently reliable for a desired application. The threshold in our graphs that follows was selected as the median of the MBD distance statistic. However, this is a user choice which will depend on the application at hand. If the algorithm is highly sensitive to noise and requires high degree of stability, then threshold will be low. For an application which is less sensitive (say one which utilizes mean values from a large ensemble and is immune to phase errors), the threshold for stability can be high.
\par
\begin{figure}[!htbp]
\centering
\fbox{\includegraphics[width=\columnwidth]{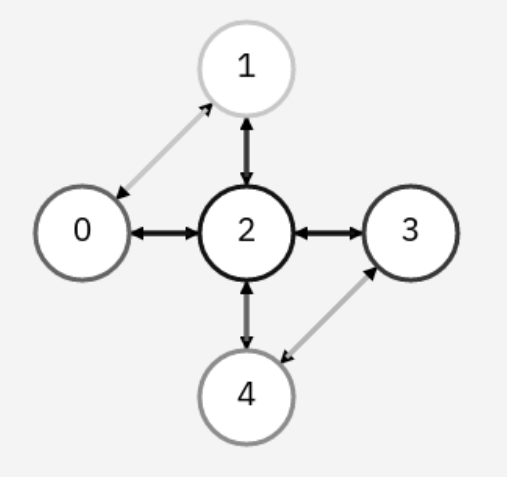}}
\caption{A physical schematic of the five-qubit transom device Yorktown from IBM with labeling of individual register elements. The circles represent qubits while the double-arrows represent connectivity for performing 2-qubit gates such as CNOT. For example, if a CNOT is needed between qubit 1 and 3, then an indirect route will be required with multiple swap gates as there is no direct connectivity between 1 and 3.}
\label{fig:yorktown}
\end{figure}

 
In this section, we develop a framework to evaluate the reliability of NISQ devices in terms of the design metric introduced in Sec.~I. Recall from Table~\ref{tab:dvz} that the first metric quantifies the capacity of the register in the NISQ device, which is a measure of the maximal amount of information that may be stored in the register. Presently, many quantum technologies use a static number of register elements and therefore we do not assess this metric. Below, we describe how the remaining four metrics are assessed. For the figures and analysis that follow, the following general comments apply unless otherwise noted:
\begin{itemize}
\item For our analysis, we collected data for a transmon device from IBM called the Yorktown. The physical layout of this five-qubit device is shown in Fig.~\ref{fig:yorktown} with the corresponding register elements labeled 0 to 4.
\item The data history spans a period of 13 months from 1 March 2019 to 31 March 2020. The statistical analysis that follows used 396 data points for each parameter. Although the data was collected twice a day at 12 hour interval, on majority of days only one unique value was seen per day as IBM has been re-calibrating only once per day. 
\item The publicly-available data provided by IBM is processed `as-is' and we defer efforts to validate the experimental data collection process itself.
\item Temporal stability assessments use March 2019 data as the reference distribution.
\item Spatial stability assessments make use of the entire temporal history from March 2019 to March 2020.
\item The horizontal, red dashed line (when present) represents the stability threshold.
\end{itemize}

\subsection{Initialization Fidelity}
\begin{figure}[!htbp]
\centering
\includegraphics[width=\columnwidth]{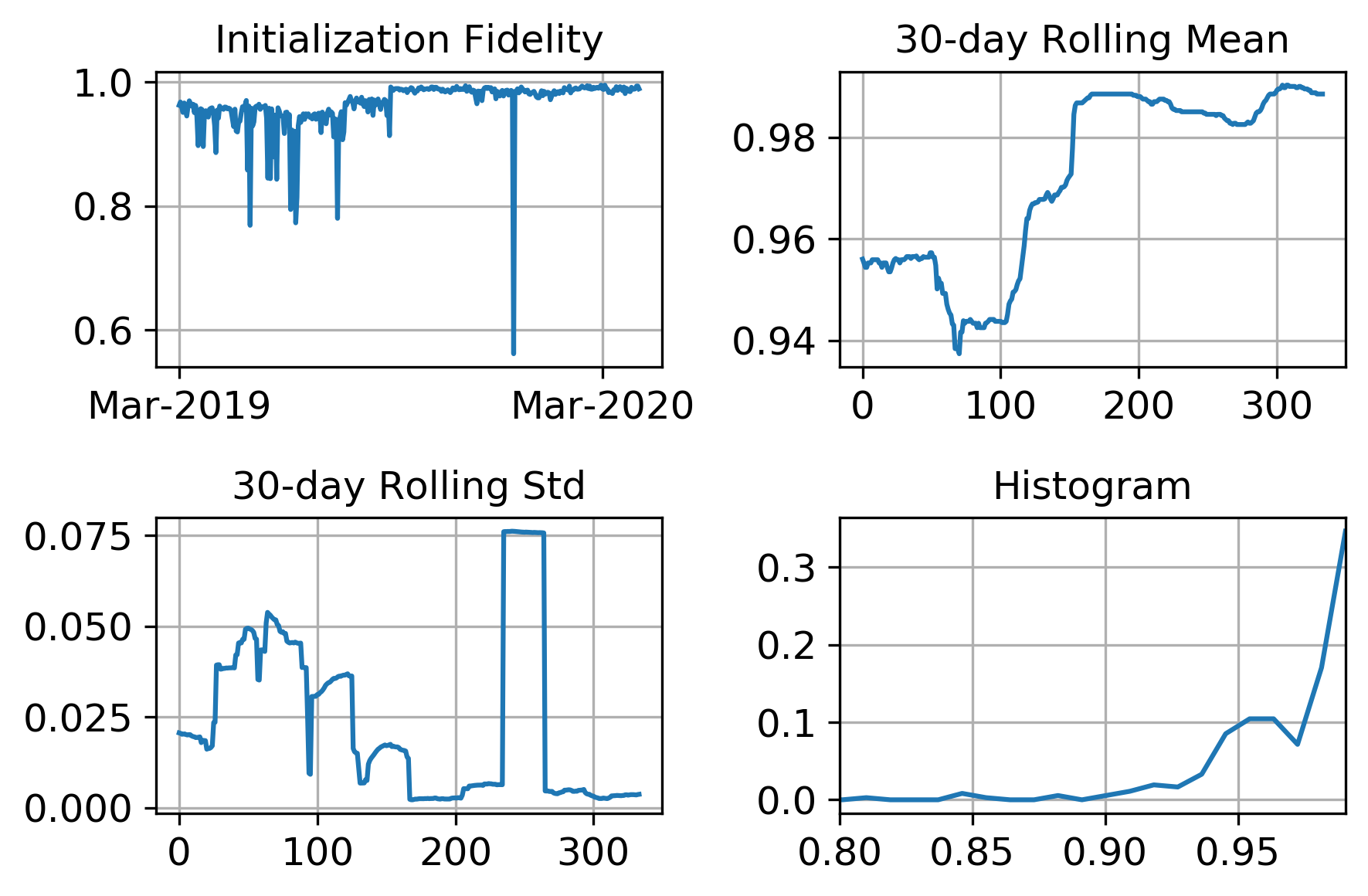}
\caption{Initialization Fidelity data for qubit 0: (a) time-series, (b) the 30-day rolling standard deviation, (c) the 30-day rolling mean, and (d) the histogram.}
\label{fig:FI_stats}
\end{figure}

Initialization plays an important role in device reliability and studying the stability of the initialization fidelity is important to systematically detect, diagnose, quantify and eventually correct time-dependent errors. The initialization fidelity $F_I$ quantifies the accuracy with which a quantum state is prepared in the NISQ device and we define it as
\begin{equation}
F_I = 1 - e_R(x, t)
\end{equation}
where $e_R(x,t)$ is the observed rate of errors when measuring the information stored at a given time $(t)$ and location $(x)$ in the device layout. An analysis of the readout error rate for the IBM Yorktown device is shown in Fig.~\ref{fig:FI_stats} in terms of the (a) time-series for Initialization Fidelity for qubit 0, (b) the 30-day rolling standard deviation, (c) the 30 day rolling mean of Initialization Fidelity, and (d) the distribution of observed Initialization Fidelity over the 13-month period. Similar analyses were completed for the other 4 qubits of Yorktown which yielded similar results.
\par
\begin{figure}[!htbp]
\centering
\includegraphics[width=\columnwidth]{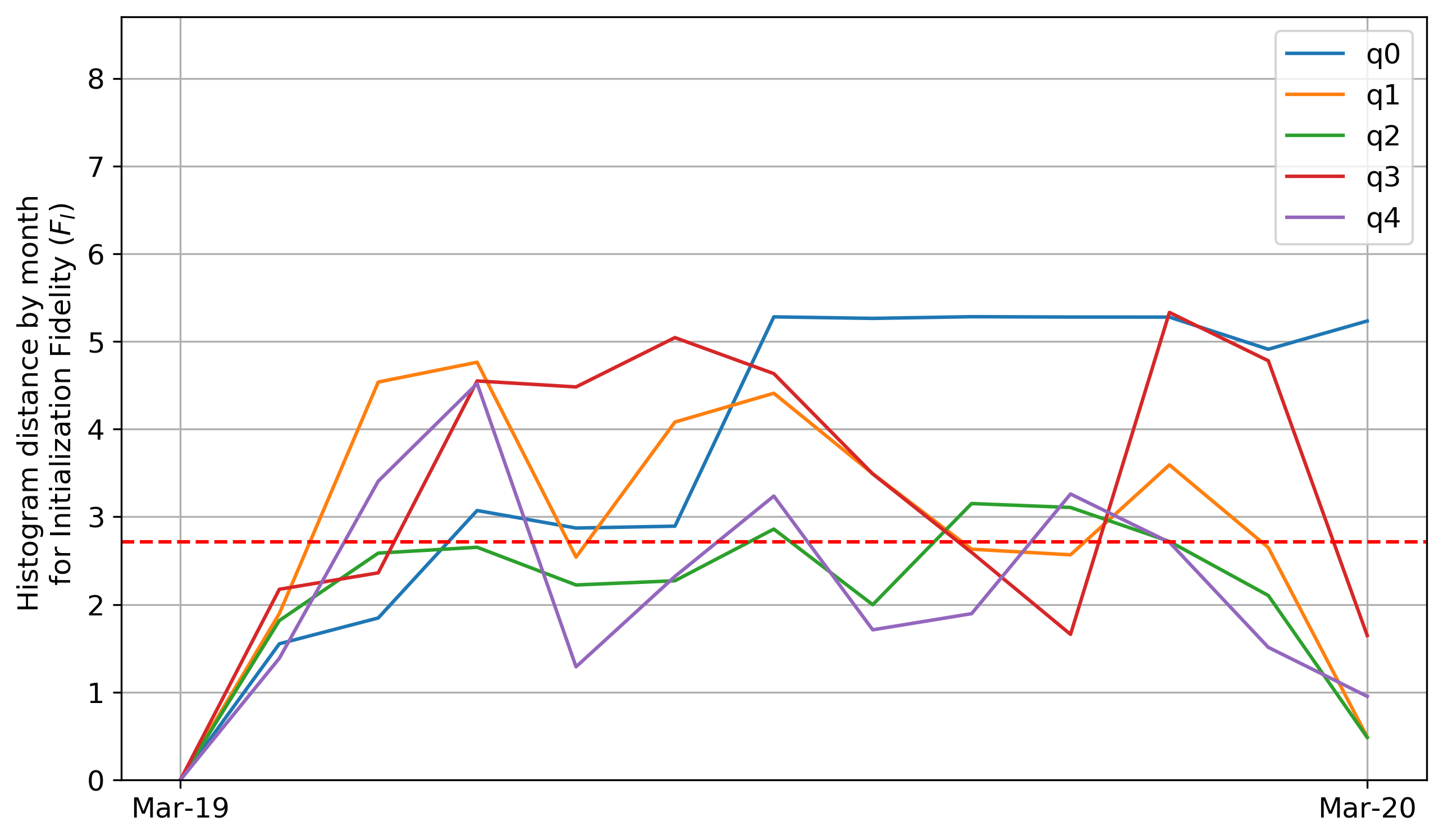}
\caption{The temporal stability of the initialization fidelity (monthly fluctuation).}
\label{fig:readout_err_control_chart}
\end{figure}

\begin{figure}[!htbp]
\centering
\includegraphics[width=\columnwidth]{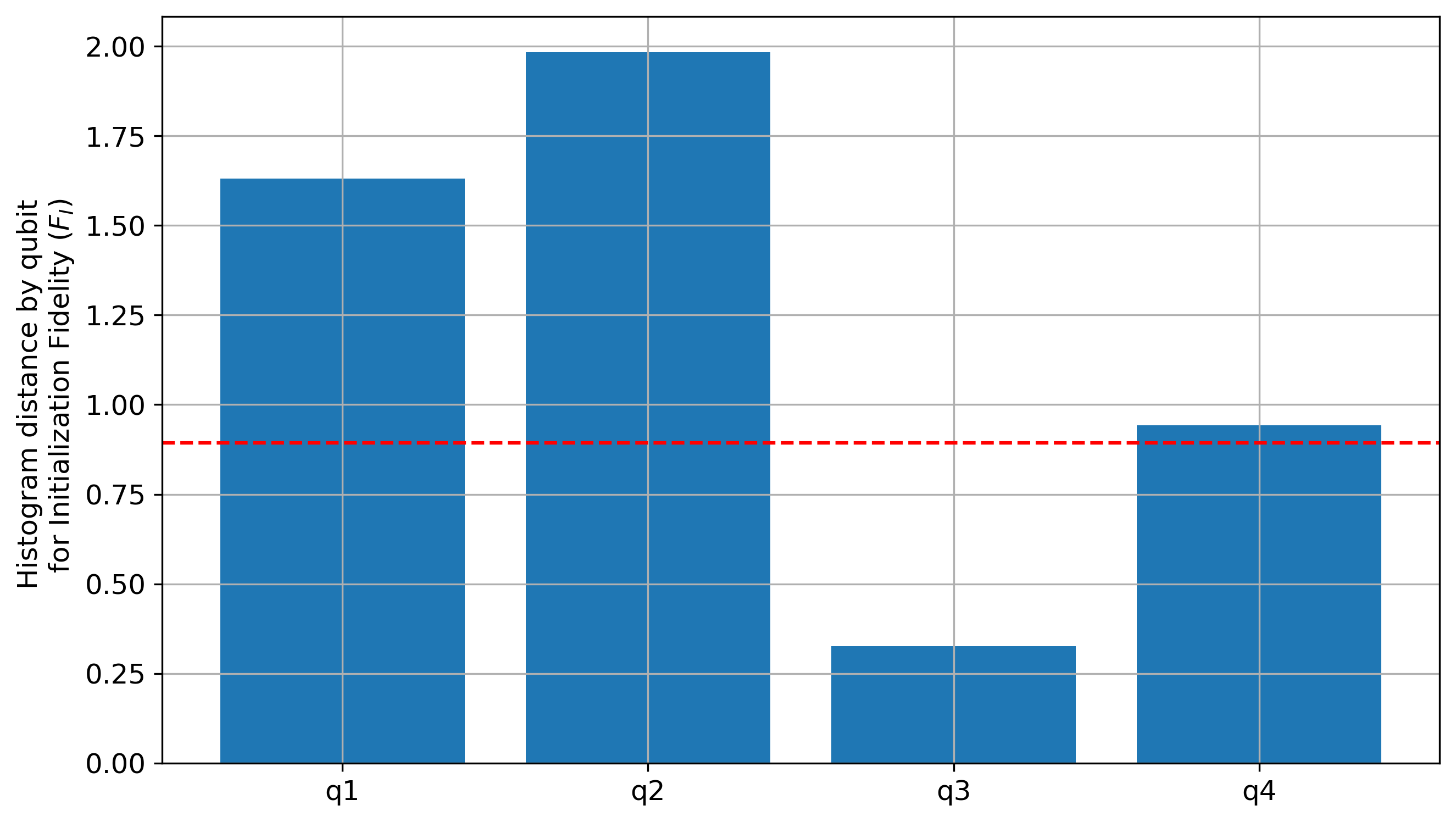}
\caption{The spatial stability of the initialization fidelity (referenced to qubit 0).}
\label{fig:FI_spatialControlChart}
\end{figure}

 
The recorded error rate data was used to generate temporal and spatial control charts that monitor the stability of the initialization fidelity in terms of the moment-based distance. In Fig.~\ref{fig:readout_err_control_chart}, we plot the moment-based distance for the initialization fidelity of each qubit with respect to time in order to assess temporal stability. The distance is measured from a reference distribution for each qubit recorded in March 2019. Therefore, the moment-based distance $d_4$ necessarily starts at zero in March 2019, but we observe growth and fluctuations over time for all 5 qubits. The horizontal dashed line provides an example of how a stability threshold may be applied to differentiate when qubits are sufficiently stable relative to the baseline distribution. A similar analysis of spatial stability is shown in Fig.~\ref{fig:FI_spatialControlChart}, which plots the moment-based distance $d_4$ the error rate data for qubit 0 (for all time) as the distance reference distribution. (To get a better sense of what is being compared, please see an example of two sample histograms in Fig.~\ref{fig:tau_spatial_twoHists.png}.)
\par

\subsection{Gate Fidelity}
Gate fidelity measures the accuracy with which a general quantum operation prepares an expected output state. A variety of methods have been developed for assessing gate fidelity, including the complete characterization of the noisy operator using process tomography. However, complete characterization is notably resource intensive and intractable as the dimension of the operator increases. Partial characterization is also possible and we will rely on partial characterization data from the IBM Yorktown device that was collected using randomized benchmarking of the 2-qubit Clifford group \cite{aleksandrowicz}. The application of randomized benchmarking relies on measuring the survival probability from a sequences of randomly selected Clifford elements \cite{magesanMay2011}. The resulting sequence of fidelity is then fit to a linear model to eliminate the influence of errors due to state preparation and measurement. From these fitted parameters, an error per Clifford gate, $\epsilon$, is deduced. In our stability analysis below, we therefore define gate fidelity as
\begin{equation}
    F_G = 1 - \epsilon_{G}
\label{eqn:fg}
\end{equation}
\par
\begin{figure}[htbp]
\centering
\includegraphics[width=\columnwidth]{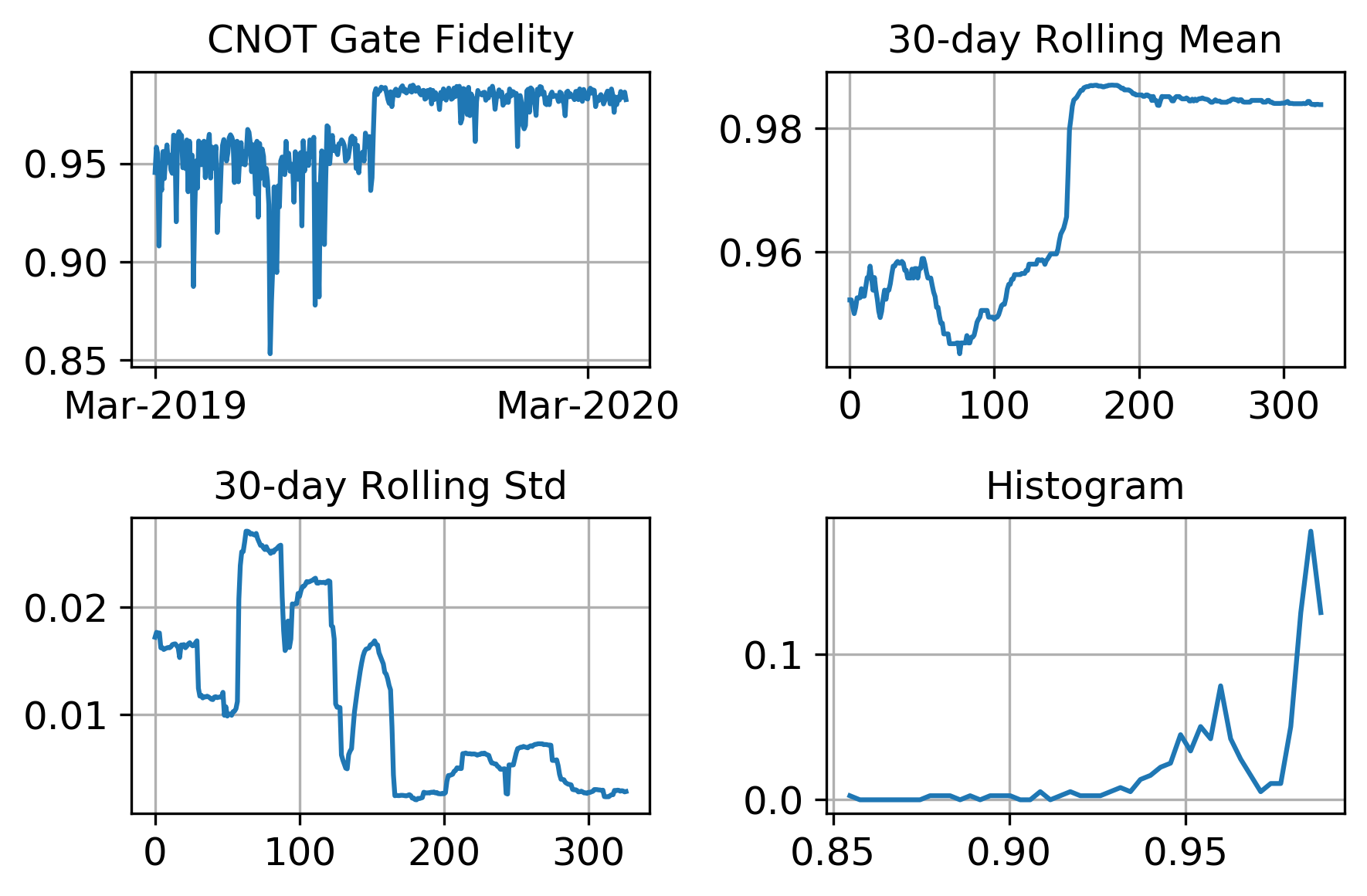}
\caption{CNOT Gate Fidelity data for qubits 0 and 1: (a) time-series, (b) the 30-day rolling standard deviation, (c) the 30-day rolling mean, and (d) the histogram.}
\label{fig:FG_stats}
\end{figure}

As an example of assessing the stability of gate fidelity, we use the CNOT gate characterized by randomized benchmarking on the IBM Yorktown over the period of March 2019 to March 2020. An analysis of the CNOT Gate Fidelity for the IBM Yorktown device is shown in Fig.~\ref{fig:FG_stats} in terms of the (a) time-series for the CNOT Gate Fidelity between qubits 0 and 1, (b) the 30-day rolling standard deviation, (c) the 30-day rolling mean of CNOT Gate Fidelity, and (d) the distribution of CNOT Gate Fidelity over the 13-month period. Similar analyses were completed for the remaining 5 possible CNOT gates in the Yorktown layout. The error value $\epsilon$ used to calculate $F_G$ in Equation~(\ref{eqn:fg}) was obtained from publicly available IBMQ data. It is seen that $F_G$ becomes almost constant towards the latter half of the data set. In this study, we have used the data as provided and have not attempted to verify or validate it.
\par
\begin{figure}[htbp]
\centering
\includegraphics[width=\columnwidth]{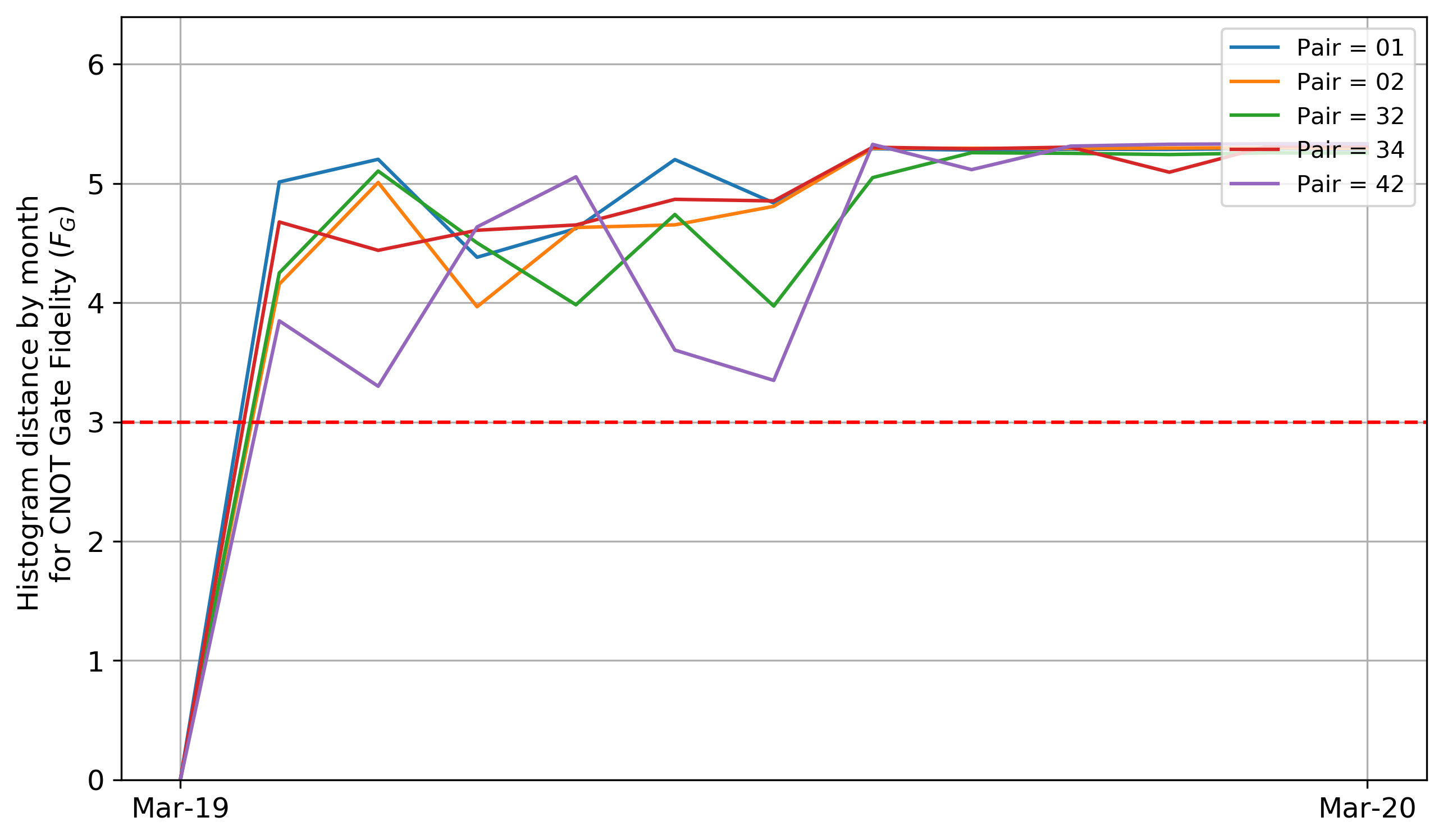}
\caption{The temporal stability of the CNOT Gate Fidelity (referenced to March 2019 data).}
\label{fig:gate_err_control_chart}
\end{figure}
\begin{figure}[htbp]
\centering
\includegraphics[width=\columnwidth]{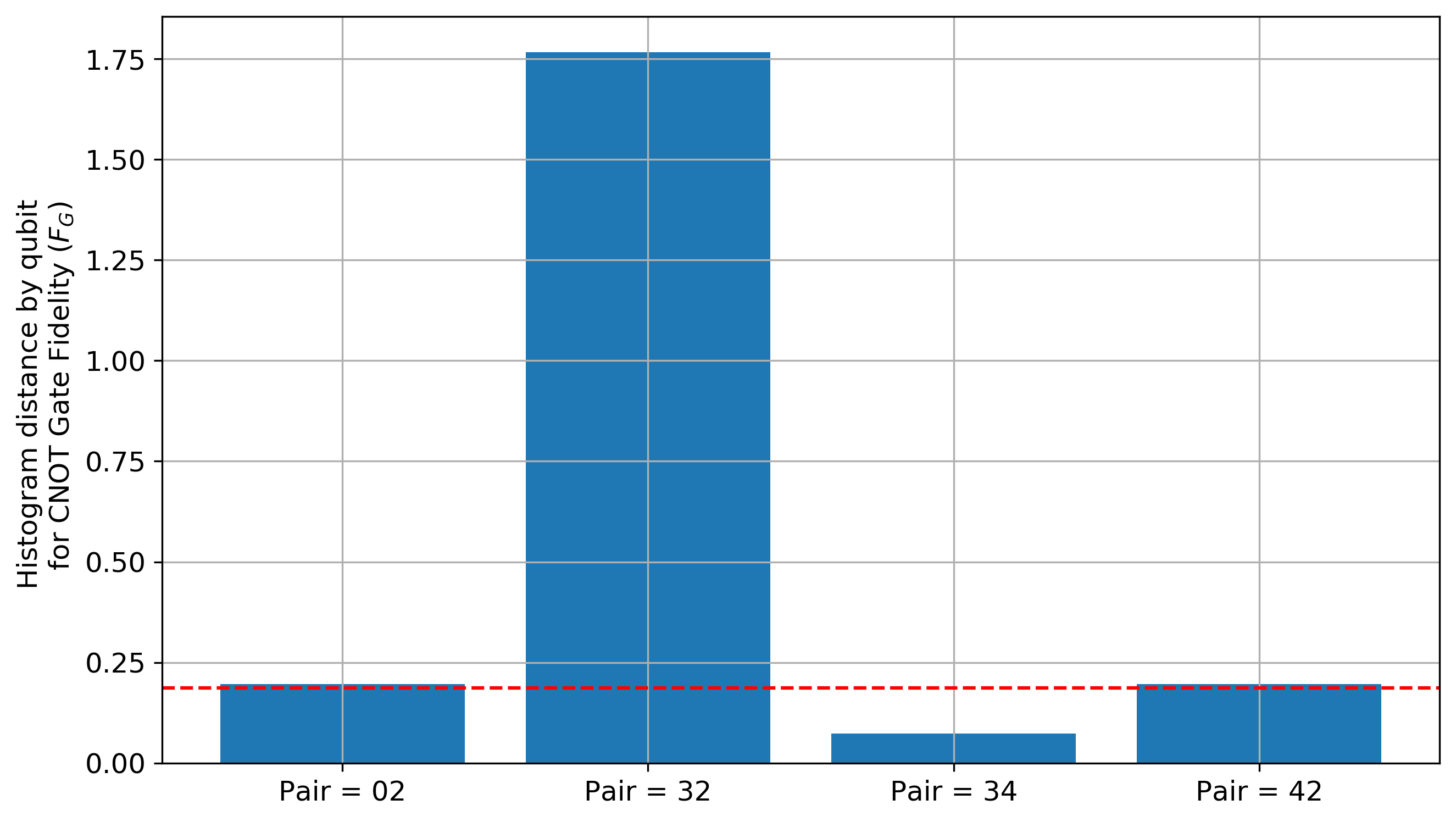}
\caption{The spatial stability of the CNOT Gate Fidelity (referenced to pair (0,1) for Mar 2019 - Mar 2020).}
\label{fig:FG_spatialControlChart}
\end{figure}

In Fig.~\ref{fig:gate_err_control_chart}, we plot the moment-based distance for the CNOT Gate Fidelity with respect to time in order to assess temporal stability. The distance is measured from the reference histogram. The reference histogram for any qubit is the distribution of the CNOT gate Fidelity for that qubit as measured in March 2019. Thus, the distance of a qubit's  CNOT gate Fidelity in March 2019 is zero. The subsequent points measure the distance between the reference distribution and the distribution as observed in a specific month. As before, the moment-based distance $d_4$ necessarily starts at zero in March 2019, and we again observe growth and fluctuations over time for all CNOT gates.  A similar analysis of spatial stability is shown in Fig.~\ref{fig:FG_spatialControlChart}, which plots the moment-based distance $d_4$ for the CNOT Gate Fidelity data with qubit pair (0,1) as the distance reference distribution. As is apparent from the stability analysis, the CNOT gate fidelity varies in both time and space.
\par
Note that when measuring gate fidelity, there is the risk of conflating with initialization fidelity. However, as mentioned before, we have used the IBM data as provided and not verified its accuracy or process rigor.
\subsection{Duty Cycle}
We quantify DiVincenzo's third criteria in terms of a duty cycle $\tau$, which is defined as the ratio of gate duration to coherence time. The coherence times set the timescales over which quantum information is lost \cite{gambetta2019} and, therefore, plays an important role in determining the number of operations that can be performed before the loss of coherence in the quantum state. Assuming a simplified model of exponential decay, and neglecting all other processes, the decoherence of a single-qubit density matrix may be characterized by the time-dependent decay of the off-diagonal coherence terms represented as
\begin{equation}
\rho(t) =
\begin{pmatrix}
\rho_{0,0} & \rho_{0,1} e^{-t/T_2} \\
\rho_{1,0} e^{-t/T_2} & \rho_{1,1}
\end{pmatrix}
\end{equation}
where the characteristic time $T_2$ defines the decoherence rate. Hence, we define the duty cycle as
\begin{equation}
    \tau = T_G/T_2
\end{equation}
with $T_G$ a characteristic gate duration for a gate $G$. We note that $T_G$ may vary with control pulse design as well as the addressed element of the register. We do not account for these variations in our analytical examples. Rather we again use the CNOT gate as a proxy for the remaining gates.
\par
\begin{figure}[htbp]
\centering
\includegraphics[width=\columnwidth]{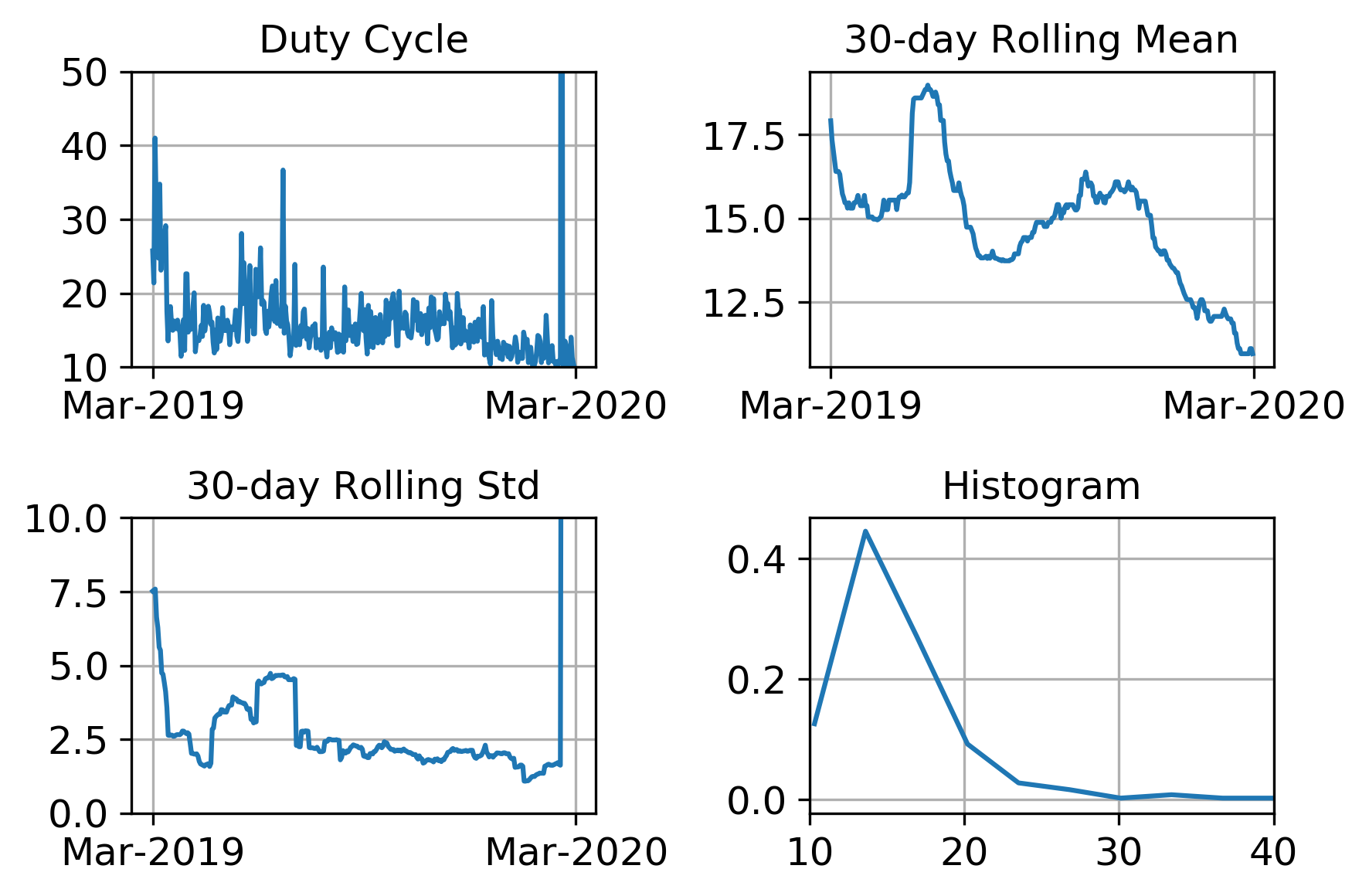}
\caption{Duty cycle for qubit 0: (a) time-series, (b) the 30-day rolling standard deviation, (c) the 30-day rolling mean, and (d) the histogram.}
\label{fig:tau_level_stats}
\end{figure}

 
As shown in Fig.~\ref{fig:tau_level_stats}, we assess stability of the duty cycle for the IBM Yorktown over the period of March 2019 to March 2020. The analysis includes the (a) time-series for the $\tau$ for qubit 0, (b) 30-day the rolling standard deviation, (c) the 30-day rolling mean of CNOT $\tau$, and (d) the distribution of observed $\tau$ over the 13-month period. Similar analyses were completed for the remaining pairs in the Yorktown layout.
\begin{figure}[htbp]
\centering
\includegraphics[width=\columnwidth]{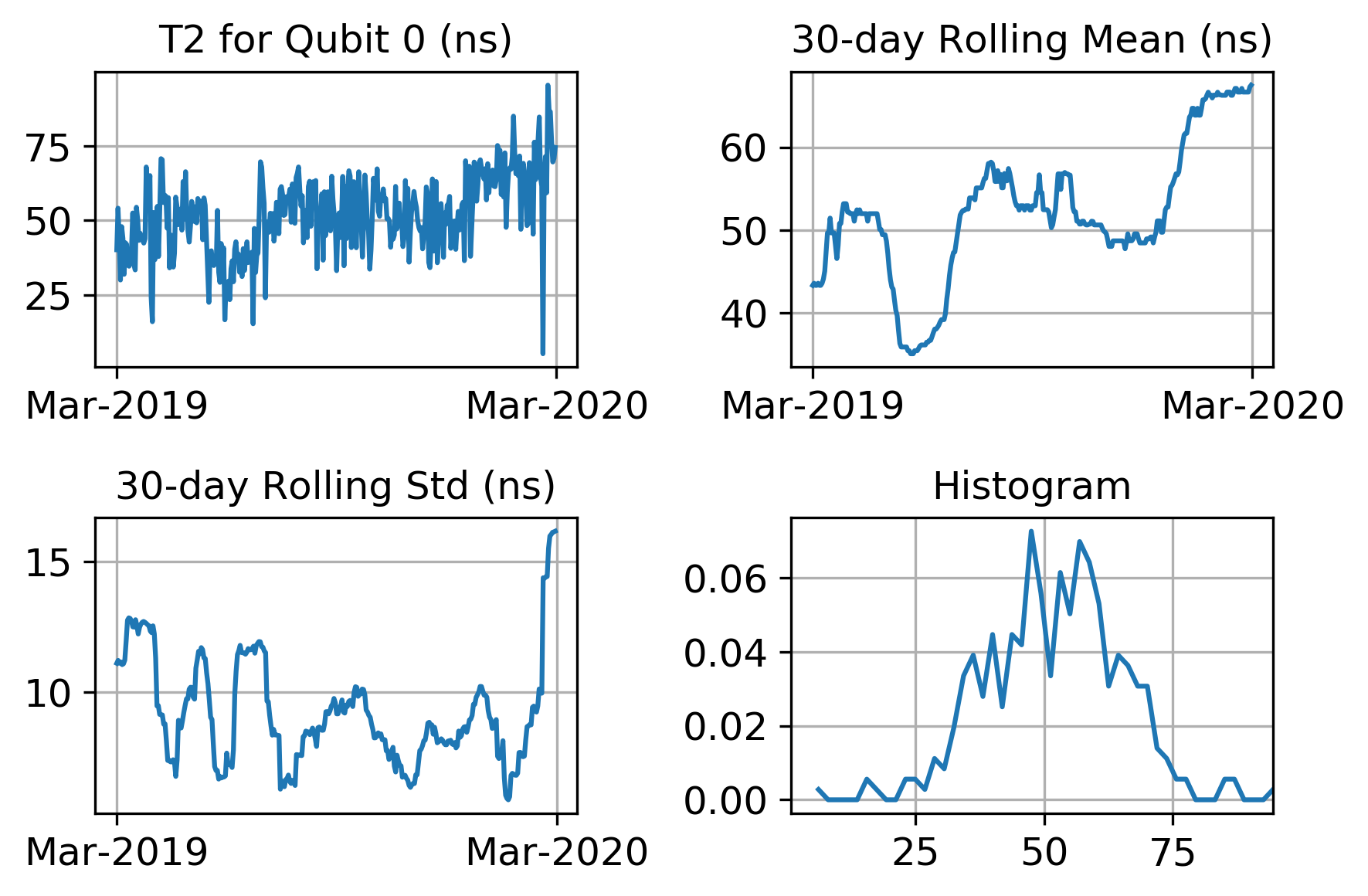}
\caption{T2 time for qubit 0.}
\label{fig:t2_level_stats_q0}
\end{figure}

\begin{figure}[htbp]
\centering
\includegraphics[width=\columnwidth]{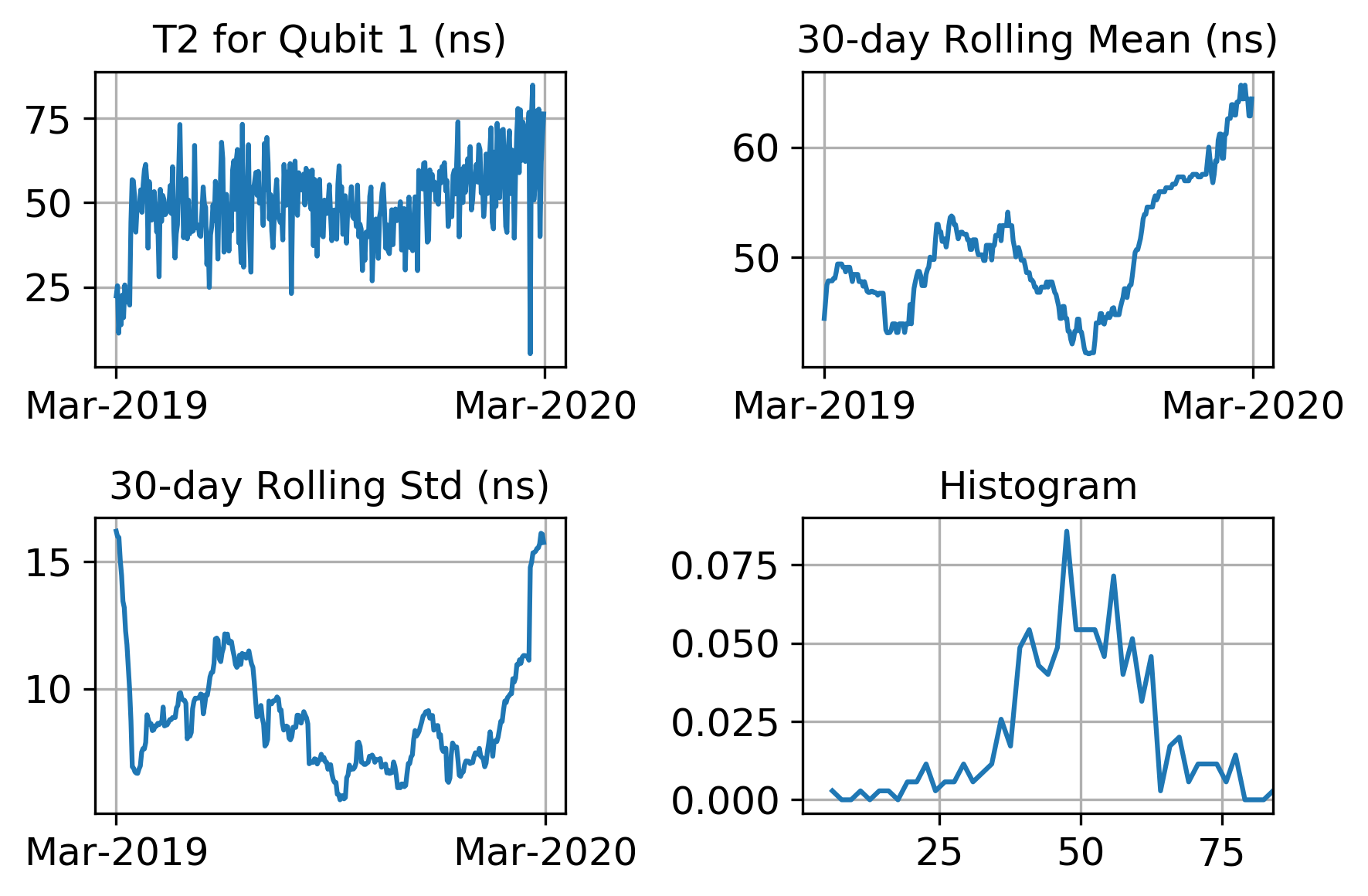}
\caption{T2 time for qubit 1. }
\label{fig:t2_level_stats_q1}
\end{figure}

\begin{figure}[htbp]
\centering
\includegraphics[width=\columnwidth]{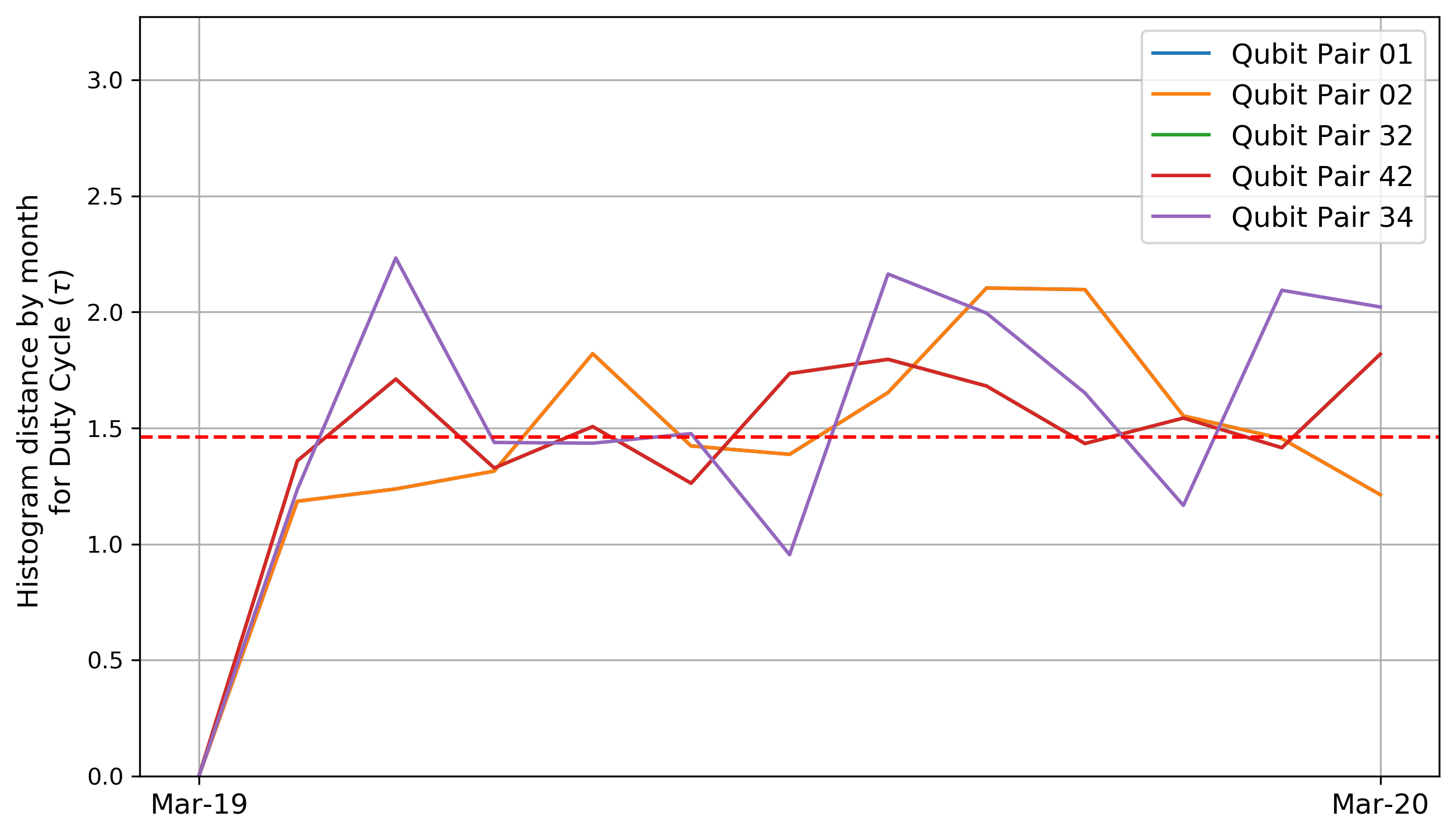}
\caption{The temporal stability of the duty cycle.}
\label{fig:tau_temporalControlChart}
\end{figure}

\begin{figure}[htbp]
\centering
\includegraphics[width=\columnwidth]{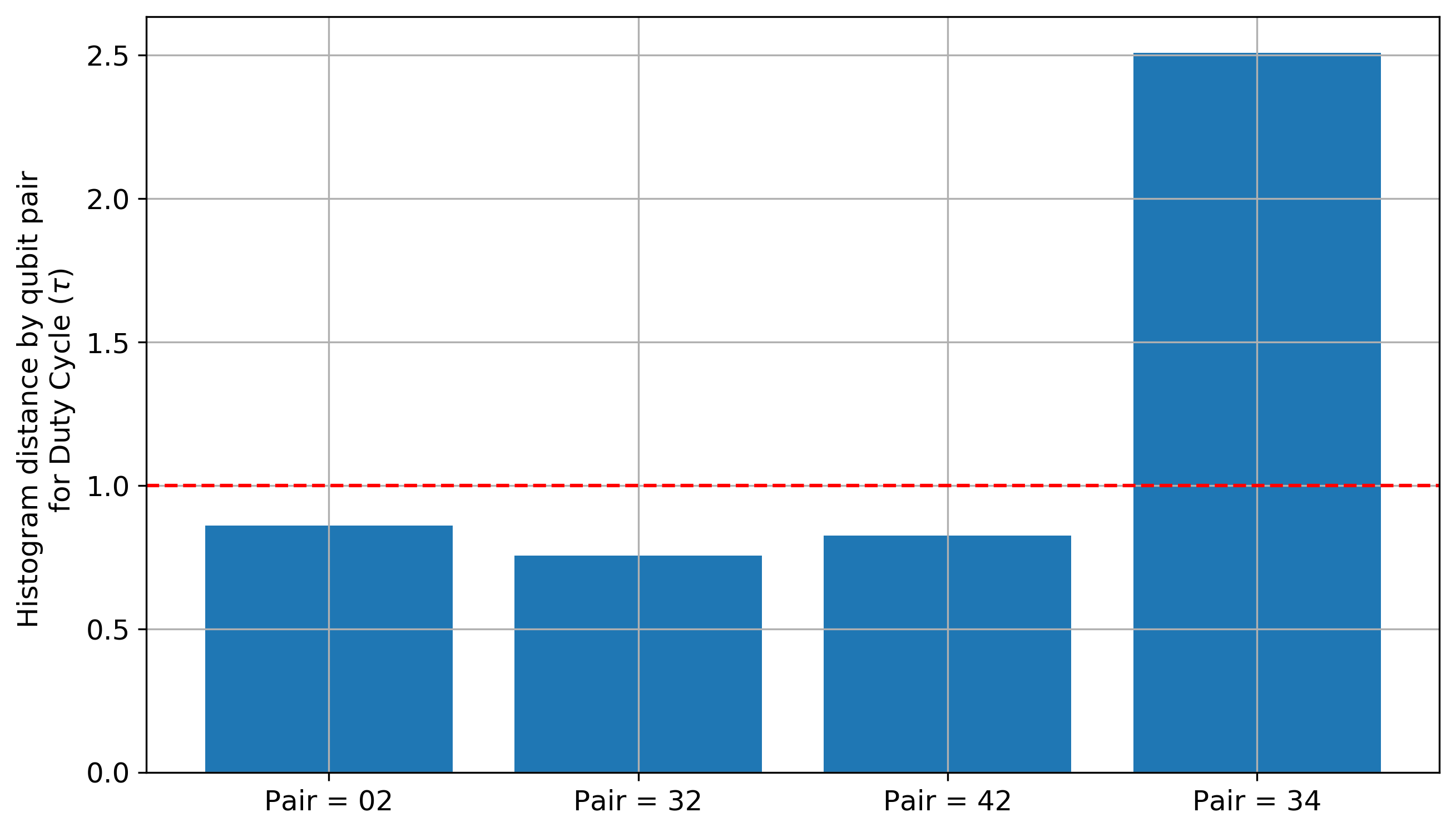}
\caption{The spatial stability of the duty cycle.}
\label{fig:tau_spatialControlChart.png}
\end{figure}

The recorded duty cycle was used to generate temporal and spatial control charts that monitor the stability of coherence time in terms of the moment-based distance. Note that the numerator of $\tau$ is fixed to the average value published by IBM. The effect of pulse duration variation can also be explored but was neglected for the present analysis. In Fig.~\ref{fig:tau_temporalControlChart}, we plot the moment-based distance for duty cycle of each qubit with respect to time in order to assess temporal stability. The distance is measured from a reference distribution for each qubit recorded in March 2019. Therefore, the moment-based distance $d_4$ necessarily starts at zero in March 2019, but we observe growth and fluctuations over time for all 5 qubits. The horizontal dashed line provides an example of how a stability threshold may be applied to differentiate when qubits are sufficiently stable relative to the baseline distribution. A similar analysis of spatial stability is shown in Fig.~\ref{fig:tau_spatialControlChart.png}, which plots the moment-based distance $d_4$ the error rate data for the qubit pair (0,1) (for all time) as the distance reference distribution.
\par
\begin{figure}[htbp]
\centering
\includegraphics[width=\columnwidth]{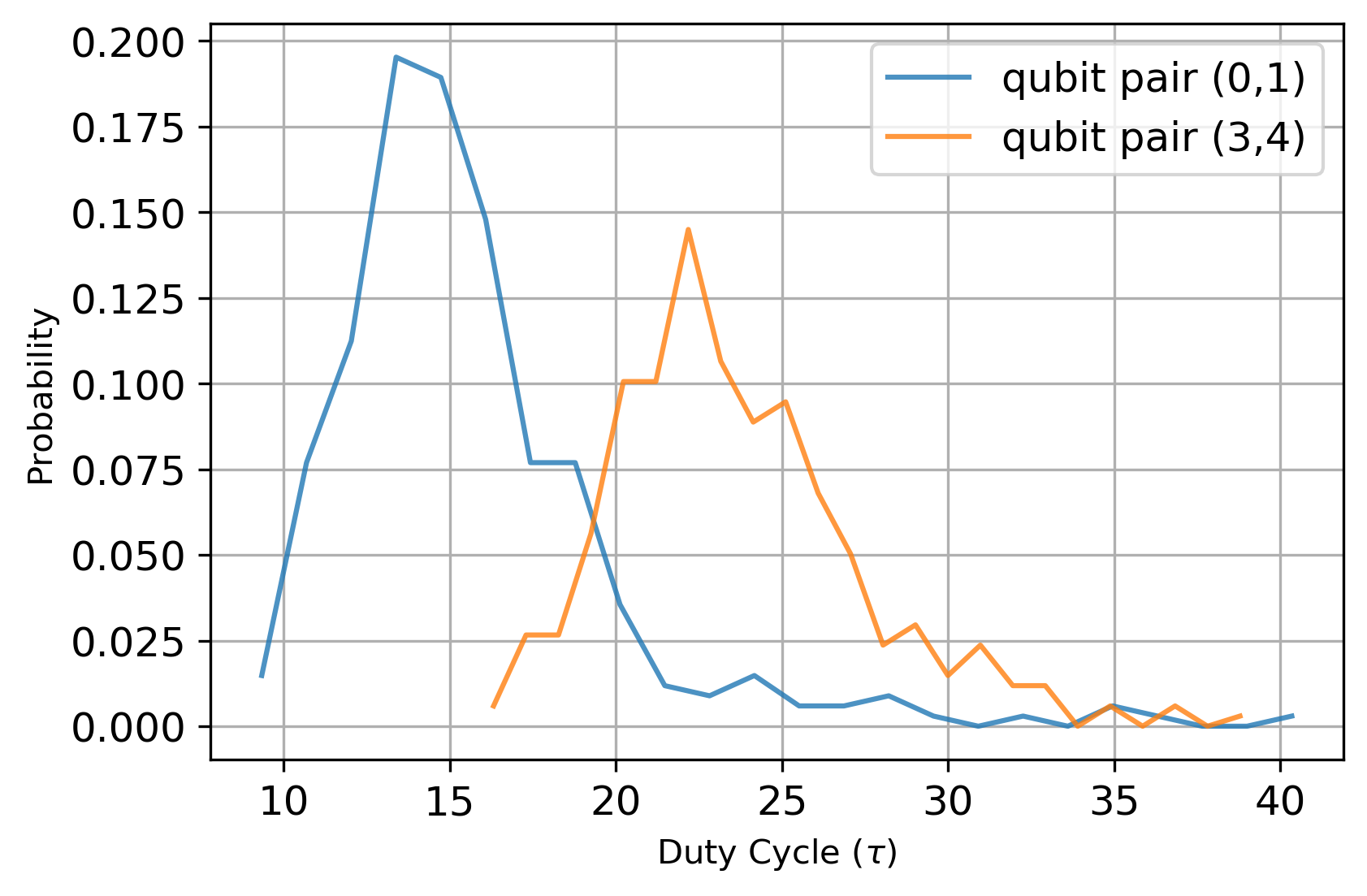}
\caption{Experimental histograms of the duty cycle $\tau$ for qubit pairs that are observed to be furthest (spatially) in the sense of the moment-based distance. The data history spans from March 2019 to March 2020.}
\label{fig:tau_spatial_twoHists.png}
\end{figure}

Figure~\ref{fig:tau_spatial_twoHists.png} shows the distributions of $\tau$ which are furthest apart in the sense of moment based distance.
\subsection{Addressability}
The Addressability $F_A$ quantifies the ability to address qubits individually. Unlike the initialization fidelity (which deals with the accuracy of the readout), this metric quantifies the ability to address qubits individually (note that the readout error can be low but addressability poor). Greater the inter-qubit correlation, worse is the single-qubit addressability. Hence, we will define Addressability in terms of the degree of correlation that arises during the measurement process.
\par
The fundamental description of an error during measurement is that the quantum state is projected into one of the computational basis states with a probability that is not given by the magnitude of the amplitude squared. Many different processes may lead to this type of erroneous projection and these processes may be broadly labeled as either correlated or uncorrelated processes. An uncorrelated error process means that the projection for an individual qubit is independent of the projection for all other qubits. By contrast, a correlated error process means that the projection for one qubit is dependent on the projection for some other qubit (1 or more). In modeling the error process, we may describe an uncorrelated projection as being separable and a correlated projection as being non-separable.
\par
For a two-qubit register, we define Addressability $F_A$ as follows:
\begin{equation}
F_A = 1-\eta
\label{eq:FM}
\end{equation}
where $\eta$ is the normalized mutual information between the two qubits:
\begin{equation}
\eta = \frac{2 I(Q_0:Q_1)}{H(Q_0) + H(Q_1)}
\end{equation}
Here, $Q_0$ and $Q_1$ represent the random binary variables describing the measurement outcomes from the first and second qubit, respectively, $H(\cdot)$ denotes the binary entropy and $I(\cdot)$ denotes the classical mutual information. Mutual information measures the amount of information obtained about one random variable by observing another random variable \cite{shannon} by determining how different the joint distribution over $(Q_0,Q_1)$ is to the product of the marginal distributions of $Q_0$ and $Q_1$. It is given by:
\begin{equation}
I(Q_0:Q_1) = H(Q_0) + H(Q_1) - H(Q_0, Q_1)
\end{equation}
where
\begin{equation}
H(Q) = -\sum_q {p(Q) \log_2 p(Q)}
\end{equation}
\par
The characterization data available for Yorktown does not have sufficient granularity to permit analysis of the mutual information or binary entropy. One needs access to the full daily  distribution of values for a period of over a year. However, that is not currently publicly available. One can diligently collect the distribution data themselves every day say for a period of one year. It is obviously practically onerous and very time-consuming. Alternately, one can reach out to IBM to special access in case they have that data stored in their servers and can share it. We have not attempted that route.
\par
Instead, we will use a numerical model to show how $F_A$ quantifies correlations in the measurement process. In this model, we characterize a two-qubit device where $q_0$ is the first qubit and $q_1$ is the second qubit.
\par
For simplicity, we consider the system state prior to measurement $\ket{\alpha}$ to be one of four states in the computational basis: $\ket{0,0}$, $\ket{0,1}$, $\ket{1,0}$ and $\ket{1,1}$. During measurement, we consider this state to be initially affected by an uncorrelated binary noise process to become $\ket{\beta}$. The transition probability from $\ket{\alpha}$ to $\ket{\beta}$ is given by 
Table~\ref{table:p_table}, in which the $(i,j)$-th element denotes the probability to prepare $\ket{j}$ when the initial state is $\ket{i}$. This binary noise model is akin to the typical symmetric binary channel used to characterize uncorrelated noisy measurement \cite{megan2019}.
\begin{table}
\centering
\caption{Transition Probabilities for Uncorrelated Noise}
\vspace{1pt}
\begin{tabular}{l |c|c|c|c|}
      & $\ket{00}$ & $\ket{01}$ & $\ket{10}$ & $\ket{11}$ \\
\hline
$\ket{00}$ & $1-p$ & $p/3$ & $p/3$ & $p/3$ \\
\hline
$\ket{01}$ & $p/3$ & $1-p$ & $p/3$ & $p/3$ \\
\hline
$\ket{10}$ & $p/3$ & $p/3$ & $1-p$ & $p/3$ \\
\hline
$\ket{11}$ & $p/3$ & $p/3$ & $p/3$ & $1-p$ \\
\hline
\end{tabular}
\label{table:p_table}
\end{table}
\par
\begin{figure}[!htbp]
\centering
\includegraphics[width=\columnwidth]{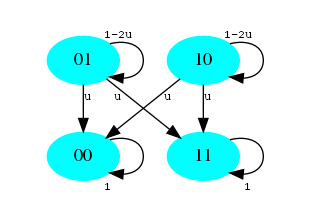}
\caption{Markov Model for the correlated error process.}
\label{fig:markov}
\end{figure}

Next we describe modeling of the correlated error process. The input to this model is the intermediate state $\ket{\beta}$, which is subjected to a correlated noise process as described by the Markov transitions matrix shown in Fig.~\ref{fig:markov}. This process prepares a final observed state $\ket{\delta}$. The originating node for each directed edge in the Markov diagram represents $\ket{\beta}$ and the destination node denotes the measured output $\ket{\delta}$. The edges contain the transition probabilities, parameterized by $u \in [0, 1/2]$. For example, $\ket{\beta} = \ket{01}$ transitions to $\ket{\delta}=\ket{00}$ with probability $u$.
\par
\begin{figure}[htbp]
\centering
\includegraphics[width=\columnwidth]{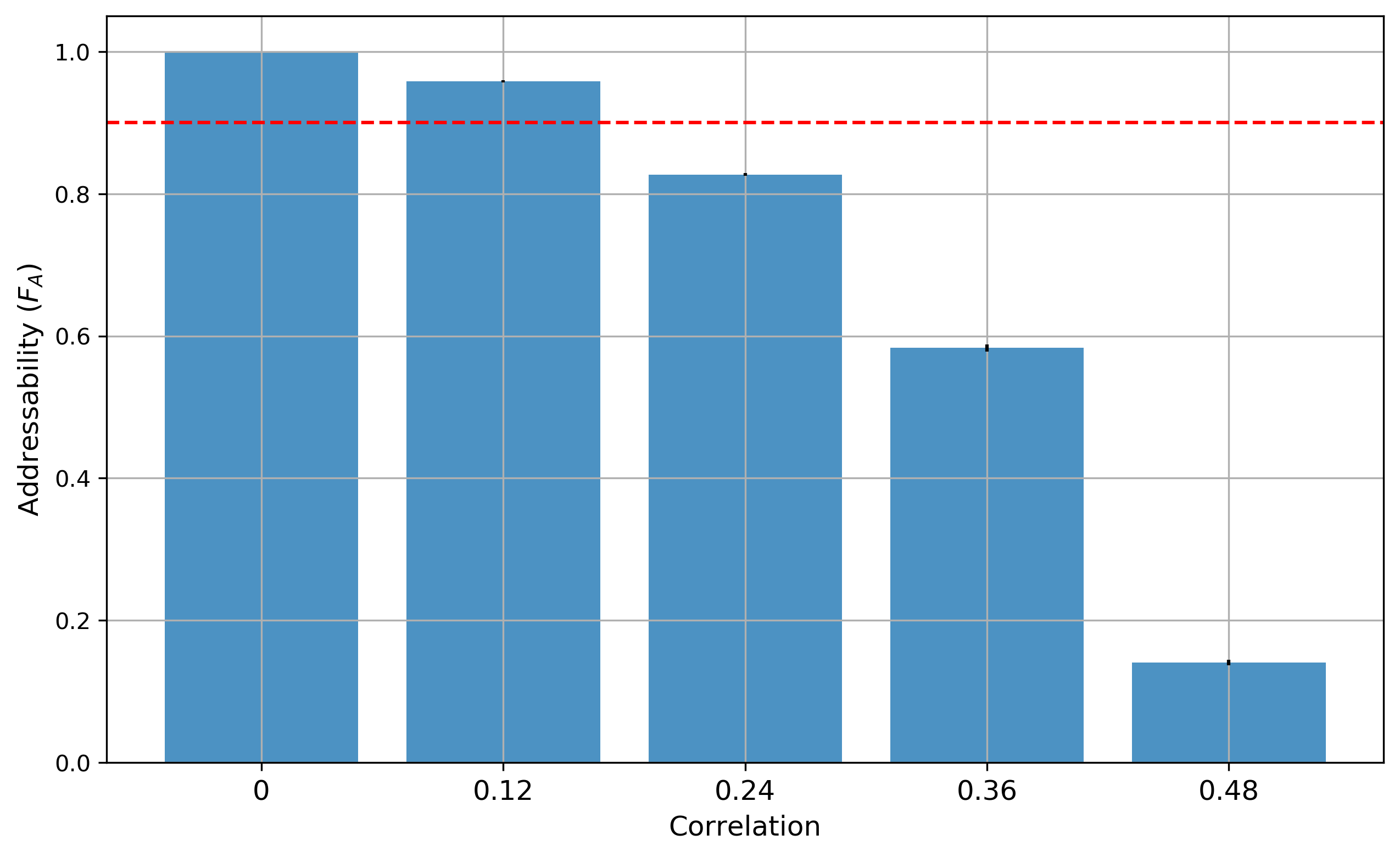}
\caption{Simulation Result for Addressability $F_A$ sensitivity to qubit correlation (u).}
\label{fig:mf_vs_p}
\end{figure}
\begin{figure}[htbp]
\centering
\includegraphics[width=\columnwidth]{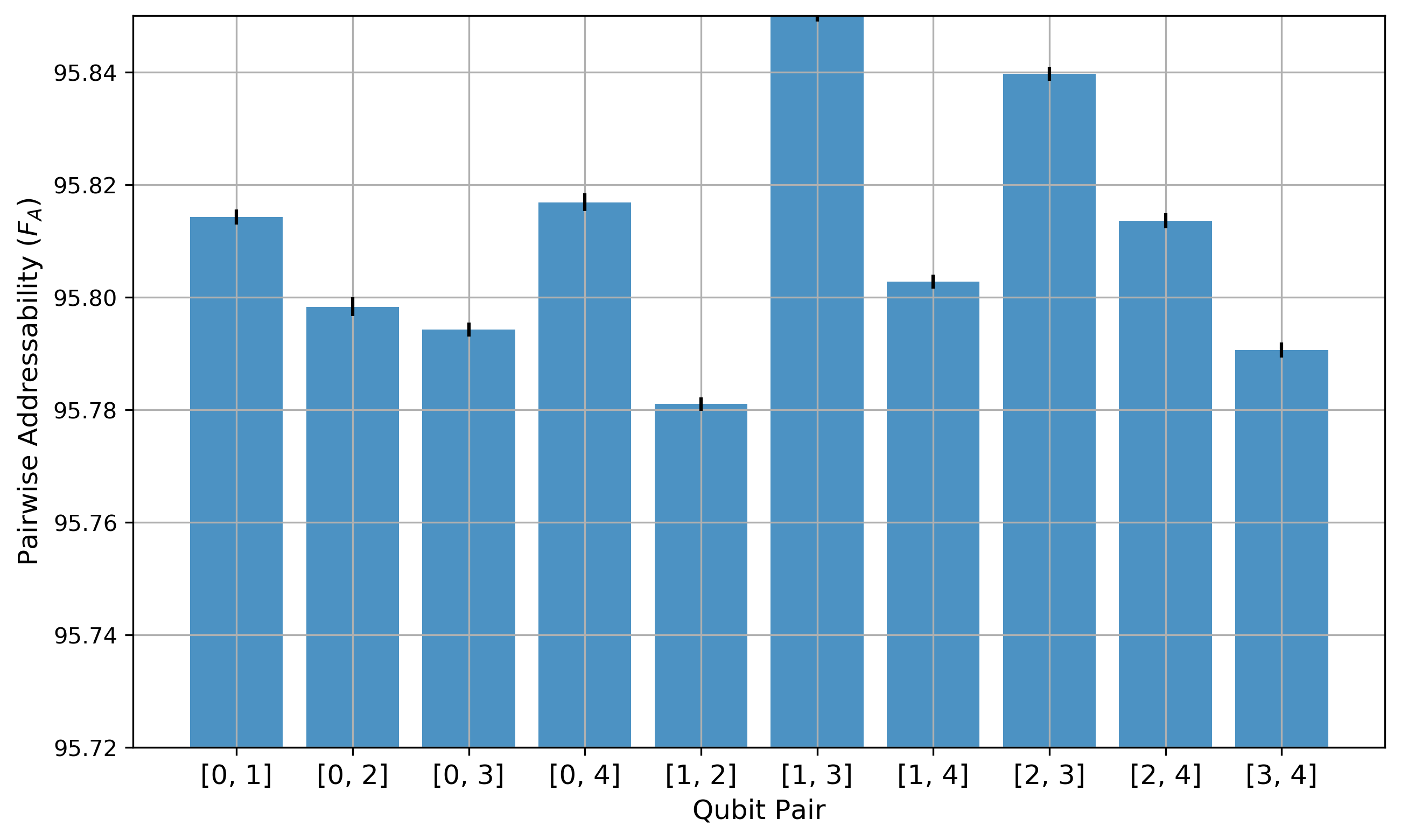}
\caption{Addressability Spatial Stability: pairwise comparison when correlation parameter $u = 0.12$. Ideally, there should have been zero spatial variation (as idealized simulations do not differentiate between qubits at different physical locations). However, small fluctuations are seen about a mean value of 95.80 with a standard deviation of 0.02. This arises due to the readout error fluctuations in the Yorktown noise model.}
\label{fig:mf_spatial}
\end{figure}

We theoretically evaluate $F_A$ for this model. 
Let $P(X)$ denote the probability of to prepare the state $X$. Using the noise models represented by Table~\ref{table:p_table} and Fig.~\ref{fig:markov}, we find
\begin{equation}
\begin{split}
P(\delta = \ket{00}) =& \frac{1+2u}{4}\\
P(\delta = \ket{01}) =& \frac{1-2u}{4}\\
P(\delta = \ket{10}) =& \frac{1-2u}{4}\\
P(\delta = \ket{11}) =& \frac{1+2u}{4}\\
\end{split}
\end{equation}
The probability of observing individual measurement outcomes $Q_0$ and $Q_1$ are therefore
\begin{equation}
\begin{split}
\textrm{Pr}(Q_0 = 0) =& 1-\textrm{Pr}(Q_0 = 1)= \textrm{Pr}(\delta=\ket{00})+\textrm{Pr}(\delta=\ket{01})\\
\textrm{Pr}(Q_1 = 0) =& 1-\textrm{Pr}(Q_1 = 1)= \textrm{Pr}(\delta=\ket{00})+\textrm{Pr}(\delta=\ket{10})\\
\end{split}
\end{equation}
which yields 
\begin{equation}
\begin{split}
\textrm{Pr}(Q_0 = 0) &= \textrm{Pr}(Q_0 = 1) = \frac{1}{2}\\
\textrm{Pr}(Q_1 = 0) &= \textrm{Pr}(Q_1 = 1) = \frac{1}{2}\\
\end{split}
\end{equation}
The binary entropy is therefore maximal, i.e., $H(Q_0) = H(Q_1) = 1$, and the mutual information is
\begin{IEEEeqnarray}{rCl}
H(Q_0, Q_1) &=& 2 - \frac{1+2u}{2}\log(1+2u)\IEEEnonumber\\
&&-\frac{1-2u}{2}\log(1-2u) \IEEEnonumber
\end{IEEEeqnarray}
This leads to a final expression for the Addressability as
\begin{IEEEeqnarray}{rCl}
F_A &=& 1 - \frac{1+2u}{2}\log(1+2u)\IEEEnonumber\\
&& - \frac{1-2u}{2}\log(1-2u) \IEEEyesnumber \label{eq:MFtoy}
\end{IEEEeqnarray}
This analysis clarifies the expectations of $F_A$ for the case of a simple, Markov noise model presented in Fig.~\ref{fig:markov}. Of course, such noise may not be present in real devices where other transitions may be possible. In absence of that information, we may rely on our definition in Eq~(\ref{eq:FM}) to experimentally compute $F_A$. For purposes of demonstration, we show in Fig.~\ref{fig:mf_vs_p} the result from a numerical simulation of a correlated noise model based on the Aer simulator from IBM Qiskit. This numerical simulator generate results of a simple prepare-and-measure circuit using the intrinsic uncorrelated readout model for the IBM Yorkown device. We then apply the correlated Markov process described above to this data to generate an example of correlated measurement errors. As shown in Eq.~(\ref{eq:MFtoy}), these numerical results match our theoretical expectations. A similar spatial characterization is shown in Fig.~\ref{fig:mf_spatial}.

\section{Conclusion}\label{conc}
We have presented a framework to evaluate the stability of NISQ devices (using moment based distance). Our contribution has identified the stability of NISQ devices as a feature of fundamental importance to current testing and evaluation. We have analyzed the stability of several different types of distributions that characterize the behavior of NISQ devices. These analyses can be extended to additional characteristics, including calibration date, optimal pulse data, and other features to develop insight into the device stability. At a first glance, it may seem that the measure of stability proposed is dependent on the reference time or reference qubit chosen. However, that is not so. The reference point simply serves as the 'origin' (to use an analogy from coordinate geometry). Depending on the reference point, the \textit{value} of MBD might be large or small. But what matters is not the \textit{absolute value of the distance} but its \textit{relative change} i.e. how badly does it fluctuate over time. If the MBD stays more-or-less constant over time, then we call the device stable.
\par
Additionally, we have presented a set of quantifiable metrics in Table~\ref{tab:dvz} based on DiVincenzo's criteria for quantum computer design. We have characterized a current QPU in terms of these metrics and we have then presented the characterization of stability in NISQ devices using a framework built on discriminating between the statistical processes that describe these empirical metrics. By defining statistics that measure changes in the statistical moments, we are able to characterize the fluctuations in distributions. 
\par
The premise of our approach is that the long-term reproducibility of results from experimental quantum computer science, and in particular those using NISQ devices, hinges on the stable and reliable performance of the computer device. Without additional efforts to make current experimental results reproducible across either time or other devices, the knowledge and insights gained from today's burgeoning field of quantum computer research may be undercut by low confidence in the reported results.
\par
Our results indicate clearly distinguishable differences in the moment-based distance for several metrics. Whether these differences are sufficient to influence the reproducibility of current NISQ applications remains to be evaluated. It also remains to be seen how do the results change were the metric evaluated on a shorter time scale, say every day, hour, or minute. Or more generally, how to make a good choice for time scale and the threshold (which will probably be application dependent) at which to call the device unstable. We plan to investigate these issues in future work.
\par
In summary, we have established a framework by which to characterize the stability of NISQ devices as an essential step in supporting the reproducibility of future experiments. 
\section*{Acknowledgments}
This work is supported by the Department of Energy (DOE) Office of Science, Early Career Research Program. SD thanks Peter Lockwood, Prakash Murali, and Megan Lilly for valuable discussions. This research used computing resources of the Oak Ridge Leadership Computing Facility, which is a DOE Office of Science User Facility supported under Contract DE-AC05-00OR22725. This manuscript has been authored by UT-Battelle, LLC under Contract No. DE-AC05-00OR22725 with the U.S. Department of Energy. The United States Government retains and the publisher, by accepting the article for publication, acknowledges that the United States Government retains a non-exclusive, paid-up, irrevocable, worldwide license to publish or reproduce the published form of this manuscript, or allow others to do so, for United States Government purposes. The Department of Energy will provide public access to these results of federally sponsored research in accordance with the DOE Public Access Plan. (http://energy.gov/downloads/doe-public-279access-plan). 

\end{document}